\documentclass[12pt,preprint]{aastex} 
 
\received{} 
\accepted{} 
\slugcomment{Revised submission to ApJ} 
\newcommand{\ms}{\mbox{m s$^{-1}~$}} 
 
\newcommand{\kms}{\mbox{km s$^{-1}~$}} 
 
\newcommand{\msun}{M$_{\odot}~$} 
\newcommand{\msune}{M$_{\odot}$} 
 
\newcommand{\mjup}{M$_{\rm JUP}~$}

\newcommand{\msini}{$M \sin i~$}

\slugcomment{Submitted to {\em The Astrophysical Journal}.}
\shortauthors{California & Carnegie Planet Search} 
\shorttitle{55 Cancri: A Planet at 5 AU} 

\begin{document} 
 
\title{A Planet at 5 AU Around 55 Cancri \altaffilmark{1}}
 
\author{
Geoffrey W. Marcy\altaffilmark{2,3}, 
R. Paul Butler\altaffilmark{4}, 
Debra A. Fischer\altaffilmark{2}, 
Greg Laughlin\altaffilmark{5},
Steven S. Vogt\altaffilmark{5},
Gregory W. Henry\altaffilmark{6},
Dimitri Pourbaix\altaffilmark{7}} 
 
\email{gmarcy@astro.berkeley.edu} 
 
\altaffiltext{1}{Based on observations obtained at Lick Observatory, which
is operated by the University of California, and on observations obtained
at the W.M. Keck Observatory, which is operated jointly by the
University of California and the California Institute of Technology.}

\altaffiltext{2}{ Department of Astronomy, University of California, 
Berkeley, CA USA 94720} 

\altaffiltext{3}{ Department of Physics and Astronomy, San Francisco 
State University, San Francisco, CA USA 94132} 
 
\altaffiltext{4}{ Department of Terrestrial Magnetism, Carnegie Institution 
of Washington, 5241 Broad Branch Road NW, Washington DC, USA 20015-1305} 
 
\altaffiltext{5}{ UCO/Lick Observatory, University of California at Santa Cruz, 
Santa Cruz, CA, USA 95064} 
 
\altaffiltext{6}{Center of Excellence in Information Systems, Tennessee
State University, Nashville, TN 37203-3401.}

\altaffiltext{7}{FNRS post-doctoral fellow, Institut d'Astronomie et d'Astrophysique, Universit\'{e}
Libre de Bruxelles, C.P. 226 Boulevard de Triomphe, B-1050 Bruxelles,
Belgium.}

\begin{abstract} 
We report precise Doppler shift measurements of 55 Cancri (G8V)
obtained from 1989 to 2002 at Lick Observatory.  The velocities reveal
evidence for an outer planetary companion to 55 Cancri orbiting at 5.5
AU.  The velocities also confirm a second, inner planet at 0.11 AU.
The outer planet is the first extrasolar planet found that orbits near
or beyond the orbit of Jupiter. It was drawn from a sample of $\sim$50 stars
observed with sufficient duration and quality to detect a giant planet
at 5 AU, implying that such planets are not rare.  The
properties of this jupiter analog may be compared directly to those of
the Jovian planets in our Solar System.  Its eccentricity is modest,
$e$=0.16, compared with $e$=0.05 for both Jupiter and Saturn.  Its
mass is at least 4.0 \mjup (\msini).  The two planets do not perturb
each other significantly.  Moreover, a third planet of sub--Jupiter
mass could easily survive in between these two known planets.  Indeed a third
periodicity remains in the velocity measurements with P = 44.3 d and
a semi--amplitude of 13 \ms.  This periodicity is caused
either by a third planet at $a$=0.24 AU or by inhomogeneities on the stellar surface 
that rotates with period 42 d.   
The planet interpretation is more likely, as the stellar surface is
quiet, exhibiting $\log(R'_{\rm HK}) = -5.0$ and brightness variations
less than 1 millimag, and any hypothetical surface inhomogeneity would have to persist
in longitude for 14 yr.  Even with all three planets, an additional planet
of terrestrial--mass could orbit stably at $\sim$1 AU. 
The star 55 Cancri is apparently a normal, middle--aged
main sequence star with a mass of 0.95 \msune, rich in heavy
elements ([Fe/H] = +0.27).  This high metallicity raises
the issue of the relationship between its age, rotation, and chromosphere.
\end{abstract}

 
\section{Introduction} 
\label{intro} 

Rarely in modern astrophysics does a nearby star attract intense
scrutiny on three observational fronts.  The main sequence star 55
Cancri (= $\rho^1$ Cnc = HD 75732 = HIP 43587 = HR 3522, G8V) has been examined for
its extreme abundances of chemical elements, its close--in orbiting
planet, and its controversial disk of dust.  These three putative
properties are plausibly linked together by the formation and
evolution of planetary systems making the system rich with
implications.

The metal--rich nature of 55 Cnc was first noticed by H.Spinrad and
B.Taylor who alerted Greenstein and Oinas (1968).  They all noted its
unusually high abundance of iron and carbon relative to that in the
Sun.  The iron lines and CN molecular absorption spectral feature were
particularly prominent in blue photographic spectra.  These results
were confirmed by Taylor (1970) and indeed, Bell and Branch (1976)
reported that carbon was yet further enhanced over iron, [C/Fe]=+0.15
.  Later spectral analyses of 55 Cnc have confirmed its high
metallicity (Cayrel de Strobel et al. 1992, 2001; Taylor 1996, Gonzalez \&
Vanture 1998, Feltzing and Gonzalez 2001) with estimates of
(logarithmic) iron abundance relative to the Sun ranging from
[Fe/H]=+0.1--0.5 .  Thus 55 Cnc is regarded as a rare ``super metal
rich'' main sequence star, but confusion still remains about the
interpretation of SMR stars (Taylor 2002, Reid 2002).

A planet was reported around 55 Cnc having an orbital period of 14.65
d, an implied orbital radius of 0.11 AU, and a minimum mass of, \msini
= 0.84 \mjup (Butler et al.~1997).  It was the fourth extrasolar
planet discovered, coming after the planets around 51 Peg, 70 Vir, and
47 UMa.  The velocity residuals to the orbital fit of 55 Cnc exhibited
a monotonic increase of 90 \ms from 1989--1995 followed by an apparent
decrease in 1996.  Butler et al.~noted that these residuals constrained
a possible second planet to have a period, $P>$8 yr, and a mass,
\msini $>$ 5 \mjup.  The decrease in the velocity residuals continued
during 1997 (Marcy \& Butler 1998), supporting the planetary
interpretation.  However, without a full orbital period nor a
Keplerian velocity curve, the possibility of stellar activity as the
cause of the residuals could not be excluded.
This star joined 51 Peg (Mayor \& Queloz 1995, Marcy et al.~1997) as
members of a growing class of planet--bearing stars that have
metallicity well above solar (Gonzalez 1998, Barnes 2001, Santos 2000,
Butler et al.~2000).

A third issue arose for 55 Cnc when Dominik et al.~(1998) presented
evidence for a Vega--like dust disk based on {\it Infrared Space
Observatory} (ISO) measurements between 25 $\mu$m and 180 $\mu$m.
They detected the photosphere at 25 $\mu$m and excesses at the higher
wavelengths.  Trilling \& Brown (1998) reported resolving the disk out
to 3.2 arcsec (40 AU) with near--infrared coronographic images.
Controversy over the disk detections arose when Jayawardhana et
al. (2000) found the submillimeter emission to be lower by a factor
of 100 than that expected from the disk reported by Trilling \& Brown.
Equally troubling were observations by the NICMOS near-infrared camera
on the {\it Hubble Space Telescope} (Schneider et al.~2001) which
imposed an upper limit on the near--infrared flux that was 10 times
lower than that reported by Trilling \& Brown.  A possible resolution
of the discrepancies was provided by Jayawardhana et al.~(2002) who
found three faint sources of sub--mm emission that were located 
near but not centered on 55 Cnc, 
implying that past detections of IR flux might have come
from background objects.  The NICMOS upper limit, the upper limit to
the sub--mm flux, and the detection of background field sources
suggest that no disk has been detected around 55 Cnc.  Indeed, Habing
et al.~(2001) discuss the non--negligible probability of spurious
detections of disks by ISO caused by fluctuations and by background
field sources.

The star 55 Cnc is also a visual binary, with a common proper motion
companion 7 magnitudes fainter (V=13, I=10.2), separated by 85 arcsec
corresponding to 1100 AU projected on the sky (Hoffleit 1982).  We
have measured the barycentric radial velocities for components A and B
to be 27.3$\pm$0.3 and 27.4$\pm$0.3 \kms respectively (Nidever et
al.~2002).  Thus, the two common proper motion stars are indeed likely
bound.  Their common space motion is similar to that of the Hyades
supercluster (Eggen 1993).

This paper will be concerned only with component A that we will refer
to as ``55 Cnc'' for which we report continued radial velocities
measurements, extending from 1989 to 2002.4.  In section 2 we provide
a update on the properties of the star, especially its mass,
metallicity, and chromospheric activity level.  In section 3 we present
all the radial velocity measurements and section 4 contains the
orbital fit to two planets.  In the remaining sections we study the
possibility of additional planets and the gravitational dynamics
between the planets.

\eject

\section{Properties of 55 Cnc}

\subsection{Stellar Surface Temperature, Metallicity and Mass}

The inferred value of the minimum mass for the planet, \msini, scales
with the two--thirds power of the adopted mass of the host star (plus
companion). Unfortunately, the mass and evolutionary status of 55 Cnc
remain uncertain despite many spectroscopic analyses.  The mass is
best derived from stellar evolution models constrained by the observed
luminosity, metallicity and effective temperature of the star.

The absolute visual magnitude of 55 Cnc is 5.47 $\pm$0.05, yielding a
luminosity of 0.61$\pm$0.04 $L_{\odot}$ from the Hipparcos parallax of
79.8 $\pm$0.84 mas (Perryman et al.~1997, ESA 1997).  Coupled with its color, B-V=0.87,
and spectral type of G8V, the star resides a few tenths of a magnitude
brighter than the zero--age main sequence.  The star's color,
spectral-type, and luminosity render it as a normal main sequence star
of modest age, a few Gyr.  If it had solar metallicity, its inferred
mass would be 0.92 \msun (Prieto and Lambert 1999).

The metallicity of 55 Cnc, however, is certainly above Solar.  Various
LTE analyses of high--resolution spectra of 55 Cnc have yielded
measured metallicities in the range [Fe/H]=0.20--0.45 (Fuhrmann et
al. 1998, Gonzalez \& Vanture 1998, Baliunas et al.~1997, Taylor
1996, Arribas \& Martinez--Roger 1989, Perrin et al.~1977, Oinas 1977,
Feltzing \& Gonzalez 2001).  These same LTE analyses yield
measurements of the effective surface temperature that fall in the
range, $T_{\rm eff}$=5100--5340 K, in agreement with the spectral type
of G8V.  Two excellent reviews of the atmospheric analyses and
interior models for 55 Cnc are provided by Ford, Rasio, and Sills
(1999) and by Henry et al.~(2000).

The above uncertainties in $T_{\rm eff}$ and [Fe/H] leave the inferred
stellar mass, derived from evolutionary models, uncertain by
$\sim$10\%.  An additional evolutionary constraint can be imposed by
age estimates, derived from the CaII H\&K chromospheric
emission, the star's position on theoretical evolutionary
tracks in the HR diagram, and from its Galactic space motion.  
The mean H\&K emission level during six
years implies an age of 4.5$\pm$1 Gyr (Donahue 1998, Henry et
al. 2000).  The age estimates from the star's placement on
evolutionary tracks range from 1 Gyr (Fuhrmann et al.~1998) to 8 Gyr
(Ford et al.~1999, Gonzalez 1998).  This lack of precision in
age-dating is disturbing and is caused primarily by the poor
atmospheric parameters of $T_{\rm eff}$ and [Fe/H].  The star is
variously suggested to be a subgiant or a Zero-age main sequence
member of the Hyades moving group (Fuhrmann et al.~1988, Eggen 1993,
Deltorn \& Kalas 2002).  The space motion velocity of 55 Cnc relative to the LSR is
29.5 \kms (Reid 2002), similar to disk stars of modest age, 2-8 Gyr.  
The modest precision of the $T_{\rm eff}$ and age of 55 Cnc suggests
that our astrophysical understanding of simple main sequence stars
leaves room for advancement.

We have independently carried out a preliminary LTE analysis for 55 Cnc based on
Keck/HIRES and Lick/Hamilton spectra at resolution 60,000.  Details
will appear in Fischer \& Valenti (2003). We find $T_{\rm
eff}$=5240$\pm$50 K, which falls in the middle of the range of
previous measurements.  Our LTE analysis yields [Fe/H]=+0.27$\pm$0.03
for 55 Cnc.  We also find that other elemental abundances are enhanced
over solar with C, Si, S, Ca, and Ni $\sim$0.3 dex above Solar
(Fischer and Valenti 2003).  Thus 55 Cnc is not only metal--rich in
iron--peak elements but even more enriched in alpha elements.

Our metallicity of [Fe/H]=+0.27 resides in
the lower half of the metallicity estimates from other groups.  
A new calibration of metallicity based on uvby photometry
has been carried out by Martell \& Laughlin (2002) and rests on 1533
calibration stars drawn from the Hauck--Mermilliod (1998) uvby catalog
and the Cayrel de Strobel (1992) spectroscopic metallicity catalog.
This calibration applied to 55 Cnc gives [Fe/H]=+0.29$\pm$0.12 and
T$_{eff}$=5220$\pm$80 
(Martell and Laughlin 2002), in agreement with our LTE spectroscopic 
analysis.  The uvby--based metallicity estimate of
Schuster \& Nissen (1989) is, [Fe/H]=0.10 +/- 0.16.

We conclude that 55 Cnc is metal--rich with [Fe/H]=+0.27 $\pm$0.10 .
Compared to stars in the solar neighborhood, 55 Cnc is metal--rich
residing $\sim$2 standard deviations from the mean metallicity (Santos
et al.~2001, Reid 2002, Butler et al.~2000).   That is,5\% of the
stars in the solar neighborhood have larger [Fe/H].   This more modest
metallicity ameliorates the difficulties in modeling the evolutionary
status of 55 Cnc (see Ford et al.~1999 and Fuhrmann et al.~1998 for
past inconsistencies).  The problem was that models predict a main
sequence for stars having [Fe/H]$\sim$+0.4 that is simply more
luminous than 55 Cnc actually is, near $T_{\rm eff}$ $\approx$5250 K.
There is no plausible evolutionary explanation for a star to reside
below the main sequence, unless in fact its metallicity is lower.
Indeed, evolutionary models with [Fe/H]=0.27 exhibit a lower (less
luminous) main sequence in the HR diagram, allowing a self--consistent
solution for 55 Cnc with a mass of 0.9-1.0 \msun and an age of 2--8
Gyr, similar to the age derived from the H\&K chromospheric emission
level and its Galactic kinematics.

Two recent mass estimates of 55 Cnc from evolutionary models yield
0.95 \msun (Ford et al.~1999) and 1.08 \msun (Fuhrmann et al.~1998).
Here, we estimate the mass of 55 Cnc from its observed luminosity,
$T_{\rm eff}$, H\&K--derived age of 5 Gyr (Henry et al.~2000), and the
new [Fe/H] value of +0.27.  We use the set of interior models provided
by Ford et al.~(1999), extrapolated from [Fe/H] of +0.39 to +0.27.  We
find a good fit to all observed parameters occurs for a mass of 0.95
\msun, the same as found by Ford et al.~(1999).  We therefore adopt a
mass for 55 Cancri of 0.95$\pm$0.1 \msun.

\subsection{Stellar Rotation}

The Ca II H\&K chromospheric emission provides two separate
determinations of the axial stellar rotation.  A predicted rotation
period can be determined from the color index (B-V) and average Ca II
flux index, $<S>$ (Noyes et al.~1984).  Henry et al.~(2000) report
$<S>$ = 0.19 during 6 yr of monitoring, implying a predicted rotation
period of 42.2 d.  We have obtained five spectra of the CaII K-line of
55 Cnc from the Keck observatory using the HIRES spectrometer (Vogt et
al.  1994).  A representative spectrum near the K--line at 393 nm is
shown in Figure 1.  The weak emission reversal in the line core is
visible to the eye, indicating that the star has a weak to modest
chromosphere and is thus relatively old (Noyes et al.~1984).  All five
Keck spectra yield $S$=0.18 within 0.02, in agreement with the value
given by Henry et al. Independently, actual periodicities have been
detected in the H\&K emission caused by magnetically active regions on
the stellar surface that rotate into and out of the visible hemisphere
(Baliunas et al.~1985).  Periods of 35--43 d have been detected for 55
Cnc (Henry et al.~2000).

This observed rotation period might vary with the phase of the
magnetic stellar cycle, as the fields migrate in latitude ala the
butterfly diagram for the Sun (Donahue, Saar, and Baliunas 1996).
 Based on the Sun, we expect rotation
periods to vary by up to 10\% on other G dwarfs as the magnetic
regions migrate in latitude.  The rotation period of 35--43 d
represents some unknown range of latitudes on the surface of 55 Cnc
during the era from 1993--2000 when the H\&K observations were made
(Henry et al.~2000).

This range for the rotation period of 55 Cnc, 35--43 d, is unlikely to
be grossly in error since it stems from actual periodicities in
emission and the range agrees with the rotation period from the mean
H\&K level (42 d).  Nonetheless, the high metallicity of 55 Cnc raises
some concern about the integrity of the Noyes calibration of rotation
period with mean S value.  Soderblom (1985) has used the Hyades stars
to test the effects of high metallicity on the standard correlations
between S value, B-V, and stellar rotation given by Noyes et
al. (1984).  The Hyades has [Fe/H] = +0.15$\pm$0.05 and many stars
have rotation rates directly measured from photometric periodicities
(Lockwood et al.~1984).  Apparently, Hyades dwarfs have longer
rotation periods (slower spin rates) by about 10\% compared to
stars of Solar metallicity (Soderblom 1985).  Thus for the metal--rich
55 Cancri, the predicted rotation period of 42.2 d may be an
underestimate of its true rotation period by 10--15\%.  If so, the
predicted rotation period would be 46--50 d, somewhat above the range
of actual observed periodicities.  Clearly a detailed study of the
calibration of activity vs. stellar rotation 
for different metallicities is needed.

Stellar rotation can also be detected photometrically.   Differential 
Str\"{o}mgren photometry with the T8 0.8~m automatic photoelectric telescope 
(APT) at Fairborn Observatory (Henry 1999) has been carried out over the past 
six years.  Figure 2 shows the results in the combined Str\"{o}mgren 
$(b+y)/2$ passband.  The photometry reveals a gradual decline in mean 
brightness of 0.002 mag.

The downward trend is undoubtedly due to long--term 
changes in the level of magnetic activity, normal for middle--aged,
inactive stars.  We have also computed a power spectrum
of the photometry.
For a brief discussion of the method see Henry et al.~(2001).
Not surprisingly, given the star's low level of activity, we find no
hint of any periodicity between 2 and 100 days.
We get similar results when we analyze each observing
season separately.  The individual seasonal light curves all show night-to-night
constancy of 0.0012 mag, the limit of photometric precision for a
single observation.

When we phase the entire data set with a prospective 44.3 day period
(see section 6) and do a least--squares sine fit, we get a
semi-amplitude of 0.00018 +/- 0.00009 mag, which constitutes a
non--detection.  Similarly, we find no signal at the 0.005 mag level
when folding the 108 Hipparcos photometric measurement of 55 Cnc at
periods near 44.3 d.

These photometric results suggest that the star is middle-aged and
chromospherically inactive.  By comparison with similar G dwarfs being
monitored for Doppler shift variations, 55 Cnc stands as a quiescent
star and is expected to exhibit velocity jitter of 3--5 \ms (Saar et al.~1998,
 Santos 2000) due to surface effects.  Moreover, no periodicities
appear, other than that of the rotation period of 35--43 d.

Rotation can also be assessed from high--resolution spectra to reveal
Doppler broadening of the absorption lines.  For 55 Cnc, estimates of
$V\sin i$ lie between 1.0 and 1.5 \kms (Soderblom 1982, Fuhrmann et
al. 1998) with uncertainties of $\sim$0.5 km/s.  For a likely radius
of the star of 0.95 R$_{\odot}$, the measured rotation period of
35--43 d implies an equatorial velocity of 1.3$\pm$0.1\kms, consistent
with the measured values of $V\sin i$.  Since the CaII H\&K rotational
periodicity is so clearly seen (Henry et al.~2000) the viewing angle
of the star cannot be nearly pole--on.  Indeed, viewing angles near
pole--on occur rarely from a statistical standpoint.  These arguments
suggest that the star is viewed within $\sim$45 deg of the equator.
Note, however, that $V\sin i$ is too poorly measured to extract $\sin
i$ directly from from known rotation period and stellar radius.

In summary, 55 Cnc is a metal--rich, middle--aged main sequence star
with a mass of 0.95 \msun.  Its chromospheric emission and photometric
variability are both low, similar to the majority of middle--aged G8
dwarfs, all of which are photospherically stable.  The stellar
characteristics of 55 Cnc are listed in Table 1.  These
characteristics all fall within the normal range found for
middle--aged G8 main sequence stars.  We expect its surface behavior,
especially photospheric velocity ``jitter'', due to surface turbulence
and spots, to be 3--5 \ms for such a star (Saar et al.~1998, Saar \&
Fischer 2000, Santos et al.~2000).  Indeed, among 1,200 FGKM dwarfs
that we are studying with precise velocities, we find the velocity jitter
to be 3$\pm$2 \ms for such middle--aged, photometrically quiet stars.

\section{Radial Velocity Observations} 
 
We have obtained 143 measurements of the velocity of 55 Cnc, during 13
years from 1989--2002.  Spectra were obtained with the cross--dispersed
echelle spectrometer, the ``Hamilton'', (Vogt 1987).  We placed a
special--purpose Iodine absorption cell in the stellar path to provide
calibration of both wavelength and the spectrometer PSF (Butler et al
1996).  The Iodine absorption cell has remained the same during the
entire 13 years, and is always temperature controlled to 50$\pm$0.1 C.
As the pyrex cell is glass-sealed, the column density of iodine
remains the same and the iodine line widths have remained constant
during the 13 years.  The dense forest of iodine absorption lines
provides an indelible record of the wavelength scale and behavior of
the spectrometer at the instant of each observation (Butler et
al. 1996).

The starlight was gathered with the 3--m ``Shane'' and 0.6--m ``CAT''
telescopes, both of which feed the Hamilton spectrometer .  The
resolution of the Hamilton was ($\lambda/\Delta \lambda$)=40,000 from
1989 to November 1994.  Starting 1994 Nov, the Hamilton camera optics
were improved by installing a new corrector plate and new field
flattener.  The new PSF has reduced wings and a more symmetric PSF.
At that same time, the wavelength coverage was expanded with a larger
CCD, from 800$^2$ to 2048$^2$ pixels.  These improvements yielded a
resolution of 55000 and higher Doppler precision by a factor of
$\sim$2.5.

The Doppler shifts of all spectra were determined by synthesizing the
composite spectrum composed of the star and iodine lines.  The free parameters in
each 2 \AA~chunk of spectrum included the linear wavelength scale,
the spectrometer PSF, and the Doppler shift.  A complete description of
our Doppler analysis is given by Butler et al.~(1996).

Velocity measurements for three arbitrary comparison stars of
spectral type G and K, similar to 55 Cnc,
are shown in Figure 3.  These stars were observed with the same
Hamilton spectrometer and telescopes at Lick Observatory
as was 55 Cnc.  The stability of the Doppler measurements
over the decade is apparent, with scatter $\sim$10 \ms or less,
and no trends over the long term.  This suggests
that the measurements carry no systematic errors greater than 10 \ms.

The velocity measurements are listed in Table 2 and shown in Figure 4.
The first 14 Doppler measurements made between 1989 and 1994 Nov have
uncertainties of typically 8--10 \ms, worse than most of the
subsequent observations due to the unrepaired optics of the Hamilton
spectrometer.  They have not been inflated by $\sim$30\% as suggested
by Cumming et al. 1999.   Observations made since 1994 Dec have uncertainties of
typically 3--5 \ms.

The standard deviation of the velocities is 69 \ms, with
peak--to--peak variations of 280 \ms.  These velocity variations are
well above the uncertainties, implying that real velocity variations
are occurring.  During time scales of a few months, the Doppler
velocity variations of 55 Cnc are dominated by a 14.65 day
periodicity, as reported by Butler et al.~(1996).  A power spectrum of
the entire set of velocities is shown in Figure 5.  The strongest peak
resides at a period of 14.65 d.  This periodicity is the same
as that reported by Butler et al.~and is caused by a planetary mass companion
having \msini = 0.84 \mjup in a nearly circular orbit ($e$=0.02)
with semimajor axis of 0.11 AU.

To test the integrity of a single--planet model for 55 Cnc, we fit
all the velocities with a simple Keplerian model, as shown in Figure 6.
The fit reveals the expected period of 14.65 d and velocity semi--amplitude of
78 \ms.  The fit is poor, with velocity residuals that exhibit an
RMS of 39 \ms, well above the errors of 3--10 \ms. Indeed, the
value of the reduced $\sqrt{\chi_{\nu}^2}$ is 10.
Clearly the single--planet model is inadequate.
Moreover, the residuals to the single--planet model
show a monotonic rise from 1989 until 1996 and a subsequent
decline (see Figure 6 in Marcy and Butler 1998).  Those residuals
exhibit coherent behavior on a time scale of a decade or more.
Indeed, the power spectrum in Figure 5 suggests a second
period near 5800 d.

\eject
\eject

\section{Models of Two Planets Orbiting 55 Cancri} 
 
We attempted to fit the velocities with a model that consisted of two
independent Keplerian orbits representing the inner planet with
$P\approx$ 14.65 d and a hypothetical outer companion.  This
two--Keplerian fit is shown in Figures 7 and 8.  In Figure 7, the
two--Keplerian fit is exhibited after subtracting the long--term
velocity variation caused by the outer companion.  The velocity
residuals have an RMS of only 12 \ms in comparison with the
single--Keplerian model (Figure 6) which had an RMS of 39 \ms.  Thus the
double--Keplerian model yielded a dramatic improvement in the quality
of the fit.  Indeed, the value of $\sqrt{\chi_{\nu}^2}$ fell to 2.7
from the value of 10 achieved with only one planet.  This
double--planet model is also shown in Figure 8 which exhibits the
observed and model velocities after subtracting the effects of the
inner planet with its 14.65 d period.  Here the wobble of the star
caused by the outer planet is revealed more clearly to the eye, though
that long--term wobble is apparent in Figure 4 as well.

The double--Keplerian model reveals an outer companion that has
period, $P$ = 14 $\pm$1.5 yr, velocity semiamplitude, $K$=45 $\pm$3
\ms, and orbital eccentricity, $e$=0.23 $\pm$0.06 .  The full set of orbital
parameters are listed in Table 3.
The uncertainties
in the orbital parameters are determined by using a Monte Carlo
simulation of the velocities, adding artificial velocity noise, and
recomputing the double--Keplerian orbital fit.
The quality of the double--Keplerian fit suggests that 55 Cnc
contains a second companion with an orbital period of
12--16 yr.

From the stellar mass of 0.95 \msun, the minimum mass of the
companion can be computed.  We find a minimum mass (\msini)
of 3.5 \mjup.  The orbital semimajor axis is 5.4 AU.

Clearly this outer companion was detected previously in the long--term
variation in the velocity residuals reported by Marcy and Butler
(1998).  But only now has enough time passed that the orbit appears to
be nearly complete.  Figure 8 shows that the velocities, after
subtracting the velocity effects of the inner 14.65 d planet, are
reaching a minimum, thus indicating the closure of the outer orbit.
This orbit closure deserves close examination.

\eject
\eject

\section{The Period of the Outer Planet}

The period of the outer planet remains somewhat uncertain
because the duration of observations, 1989--2002,
is only 13 yr, very close to the best--fit period,
14 yr.  The most recent season of velocity measurements, during 2002,
shows a flattening of the downward slope that had characterized
the velocity residuals to the single--planet fit from 1996--2001.
This flattening of the velocities is visible to the eye in Figure 8.

Nonetheless the flattening is modest, leaving 
the period of the outer planet weakly constrained.
The weak constraint on the period of the outer planet can be illustrated by
arbitrarily adopting a period of 20 yr for it, instead of the
best--fit period, 14 yr.  We recomputed the double--Keplerian fit to the
velocities but froze the value of the period of the outer planet to be
20 yr.  All other orbital parameters for both planets were allowed to float.
This constrained fit tests the viability of an orbital period of 20 yr for the outer planet.
The resulting $\sqrt{\chi_\nu^2}$ is 2.92 which is slightly worse
than the best--fit value, 2.76.  Apparently a 20 yr
orbital period is poorer than the best--fit 14 yr period for
the outer planet.

However the value of $\chi_\nu^2$ for $P$=20 yr is only slightly worse
than that for the best fit because of dilution caused by the
preponderance of measurements during 10 years prior to the recent
season.  Adopting $P$=20 yr yields a poor fit only during the past
season.  Indeed during the
past two seasons, 2001 and 2002, the velocities appear plausibly
fit by the theoretical curve for $P$=20 yr which continues a downward
slope.  Thus the $\chi^2$ statistic is not a sensitive diagnostic of
recent changes in the slope of the velocities.

To improve the sensitivity to the period of the outer planet, we
performed a different test of the recent velocities to sense a
flattening of the slope.  We fit the velocities with a single
Keplerian (with $P \approx$ 14.65 d) and a free ``trend'' in the velocities during two time
intervals, 1995.5--2001.5 (six years) and 2000.5--present (two years).
The trend during those two intervals was found to be -16.25 $\pm$0.3
\ms per year and -4.5 $\pm$ 1.8 \ms per year.  Thus, the slope
flattened during the past two years from its previous decline rate of
16 to 4.5 \ms per year.  The two slopes differ by 6$\sigma$.
Isolating just the past season alone (2001.8--2002.5) the best--fit
slope is now positive, +3.4 \ms per year.  

Thus the recent stellar velocities induced by the outer companion are
apparently flattening, and indeed appear to be increasing.  This
velocity turn--around is consistent with the best--fit
double--keplerian that yields a period of 14 yr for the outer
planet.  Nonetheless, the reality of the flattening remains only
6$\sigma$, marginal enough that further intense velocity measurements
are warranted.  Until then, the period of the outer planet remains
poorly constrained and could be as large as 20 yr at the 6$\sigma$
level.

\subsection{Orbital Eccentricity of the Outer Planet} 

The orbital eccentricity of the outer planet is formally found to be
0.23$\pm$0.06, from the two--Keplerian model.  (In section 6, the
superior three--planet model yields an eccentricity of $e$=0.16
$\pm$0.06 for the outer planet.)  The uncertainty of 0.06 is the
1$\sigma$ standard deviation of 100 Monte Carlo trials in which
artificial noise was added to the velocities and the double--Keplerian
was recomputed. Clearly the eccentricity is not constrained well, but
is probably less than 0.3 .

Our experience is that velocity errors tend to artificially increase the
best--fit eccentricity.  This artificial increase
occurs for at least two reasons.  First, the eccentricity cannot be
found negative, implying the velocity noise tends to push the
best--fit eccentricity away from $e$=0 in only the positive direction.
Second, the flexibility of the shapes of theoretical Keplerian
velocity curves allows the best--fit model to be contorted in order to
fit the discrepant velocity measurements.  
Simmilar issues of the systematic errors and distribution
of errors  are also discussed by Halbwachs et al. (2000).
This Keplerian contortion
act is especially insidious for cases in which barely one full orbit
has been completed such as this case for the outer planet.  The
eccentricity is allowed to take on a distorted value to best fit the
most recent velocities, bending the velocity curve with a large second
derivative.  The recent velocity measurements in 2002 are well fit
only by a significant curvature in the orbital fit, as seen in Fig 8.
Thus we suspect that the best--fit eccentricity of 0.23 may be 
overestimated by up to 25\%.  More careful Monte Carlo tests could be
applied to ascertain this bias.  Nonetheless, this two--Keplerian fit
is clearly not adequate and the best estimate of orbital parameters
likely comes from including a treatment of the periodicity in the
residuals.

\eject
 
\section{Velocity Periodicity of 44 days} 

The double--Keplerian fit, while a great improvement over a
single--Keplerian model, yields residuals with RMS=12 \ms and
$\sqrt{\chi^2_{\nu}}$=2.7.  Normal, chromospherically quiet stars
yield RMS = 4--6 \ms (partly due to velocity jitter) and yield values
of $\sqrt{\chi^2_{\nu}}$ less than 1.5 .  The uncertainties for each
velocity measurement are the weighted uncertainty in the mean velocity
of the $\sim$400 individual 2 \AA~spectral chunks from each spectrum.
They represent internal errors and closely match the actual
uncertainties of the velocities as shown by Cumming et al.~(1999).
Thus the observed scatter in the residuals to the double--Keplerian
fit of 12 \ms significantly exceeds the errors of $\sim$4 \ms.  The
two--Keplerian fit fails to explain the data.  We have similarly
attempted full N--body simulations of the three-body system (star and
two planets).  These calculations yield no better fits.

A power spectrum of the velocity residuals to the double--Keplerian
fit is shown in Figure 9.  The power spectrum reveals a strong
periodicity at P = 44.3 d.  This periodicity in the velocities appears
compelling even from the velocity measurements obtained within just
one season, such as in 1996 and 2002.  In principle, this 44.3 d
periodicity could be caused by some stellar phenomenon or by a third
planetary companion.

We tried three--Keplerian models to fit the velocities of 55 Cnc, as
described in section 6.1.  Such models yield $\sqrt{\chi_{\nu}^2}$ =
1.8, significantly superior to that achieved with only two Keplerian
orbits ($\sqrt{\chi_{\nu}^2}$ = 2.7).  We normally find
$\sqrt{\chi_{\nu}^2}$ = 1.1--1.5 for adequate fits to
chromospherically quiet stars.  Thus the evidence for a third planet
is plausible, but it leaves some velocity scatter unaccounted for.

Moreover, the stellar rotation period of 35--42 d is suspiciously
similar to this velocity period of 44.3 d.  The stellar rotation
period, as described in section 2.2, stems directly from observed
periodicities in the Ca II chromospheric diagnostic.  The period of CaII
emission variations represents the stellar rotation of the latitudes at
which the active regions reside. However, during a stellar cycle the active
regions probably migrate to other latitudes where differential rotation
would yield a different rotation period.  Thus, the velocity
periodicity at P = 44.5 d could, in principle, be caused by surface inhomogeneities
(spots, plage, turbulent domains) at a some latitude where the
rotation period is $\sim$44 d.

It is difficult to support the notion that surface effects cause the
44 d velocity periodicity in 55 Cnc.  The weak chromospheric of 55 Cnc
and the lack of photometric variability render its stellar atmosphere
stable and void of large velocity excursions, as described in
section 2.  The velocity jitter is expected to be less than 5 \ms for
such an old, quiet G8V star.  The observed velocity semi--amplitude of
13 \ms, found in the three--Keplerlian fit, is much larger than that
expected from surface effects.  Thus, a planet explanation is to be
favored, with all due concern about the near--coincidence between the
rotation and orbital periods of 42 and 44 d, respectively.

\subsection{Interpreting the 44-day Period as a Third Planet} 

The removal of the 14.65 d and 14 yr Keplerian periodicities from the
radial velocity data set leaves a large residual scatter with power
concentrated at a period of 44.3 day.  This velocity signal has
a semi--amplitude of 13 \ms (best--fit sinusoid), larger than expected from a
chromospherically quiet G8 main sequence star.  Indeed, quiet G8 dwarfs
have never revealed any intrinsic periodicity at measurements levels
above 5 \ms and the velocity jitter (RMS) is only
3--5 \ms for such a star (Saar et al.~1998, Santos et al.~2000)
and is observed to be stochastic on yearly time scales.
Thus surface effects are expected to cause velocities of lower
amplitude and void of coherence, quite different from the
amplitude and coherence in the 44.3 d velocity periodicity we observe.
The poor plausibility of stellar surface effects in explaining the
44.3 d period motivates
a detailed study of three--planet models here.

The addition of a third planet to the planetary system model results
in a stellar reflex motion formed from the sum of three simultaneous
Keplerian orbits. The resulting orbital parameters from this best--fit
model are listed in Table 3 and the fit is shown in Figure 10.  The
fit for a triple Keplerian model gives $\sqrt{\chi_{\nu}^2}$=1.8, an
improvement over the 2--planet fit which yielded
$\sqrt{\chi_{\nu}^2}$=2.7.  This shows that the 3--planet model of the
planetary system is plausible. This model contains inner and outer
planets not very different from the best two--Keplerian model, but
also contains a middle planet characterized by: $P$=44.3 d, $K$=13.0
\ms, $e$=0.34, implying \msini = 0.21 \mjup (see Table 3).

The $\sqrt{\chi^{2}}$=1.80 statistic for the foregoing 3-planet model
represents a considerable improvement over the 2-planet fits to the
data. Nevertheless, the 8.5 \ms velocity scatter of the 3-planet fit
would require a stellar jitter of $\sim6.5$ \ms, whereas 55 Cancri
likely has a more quiescent jitter of order 3-5 m/s, as expected for a
chromospherically quiet main sequence star (Saar et al 1998, Santos et
al 2000).

One possible explanation for the high $\sqrt{\chi^{2}}$ value in the
3-planet model stems from the assumption that the orbits are
unperturbed Keplerian ellipses. The system listed in the previous section
has a period ratio $P_{c}/P_{b}$=3.02, which is close to the 3:1
commensurability. This means that over the $\sim$5000 day timespan
that the star has been observed, the planet--planet interactions
between the inner and middle planets will tend to add in a
constructive way. This is illustrated in Figure 11, which shows the
running difference between the radial velocity curve of the star under
the influence of summed Kepler motions, and under full four-body
motion.  The discrepancy between the two versions of the stellar
motion grows to $> 100 {\rm m/sec}$ after 10 years of observation,
which indicates that self-consistent fitting (as described by Laughlin
\& Chambers 2001 and Rivera \& Lissauer 2002) is required in order to
to correctly describe a system in this configuration.

We first used a Levenberg-Marquardt minimization routine (Marquardt 1963; 
Press et al
1992) driving a four--body integrator to produce a self-consistent fit
to the radial velocity data.  As with other methods based on the
method of steepest descent, the Levenberg-Marquardt routine requires a
good initial guess in order to converge to a global minimum in the
parameter space.

The summed triple--Keplerian fit to the data provides a natural initial
guess, just as a dual--Keplerian initial guess allows the Levenberg-Marquardt
method to successfully refine the model of the 
GJ 876 system (Laughlin \& Chambers 2001; Rivera \& Lissauer 2001).
In that system, the best dual--Keplerian fit (Marcy et al 2001) has $\sqrt{\chi^{2}}=1.88$.
When this dual-Kepler fit is used as a starting condition, the 
Levenberg-Marquardt method rapidly converges to a fit with $\sqrt{\chi^{2}}$=1.55.

Unfortunately, however, the 3-Keplerian fit to the 55 Cancri data does
not provide a similarly propitious point of departure for dramatic
improvement.  When the summed Keplerian fit is inserted as an initial
guess, the code converges to a self-consistent solution with
$\sqrt{\chi^{2}}=1.85$. This fit is shown in the first column of Table
4. Experimentation with the starting conditions shows that there is a
very strong sensitivity of $\chi^{2}$ to small variations in the
initial conditions, coupled with a $\chi^{2}$ landscape that is
topologically rugged on large scales. It is therefore useful to test
additional methods in an attempt to locate the global minimum and thus
find the true configuration of the system.

We first used a scheme which turns on the planet-planet perturbations
in a gradual way, and which was successfully adopted for GJ876 by
Rivera \& Lissauer (2001). We decreased both the masses of the planets
and the magnitudes of the stellar reflex velocities by a factor of one
million. This allows the Levenberg-Marquardt N-body code to recover
the 3-Keplerian fit of the previous section.  We then gradually
increased both the masses of the planets and the radial velocities in
a series of discrete increments. After each increment, we allowed the
Levenberg-Marquardt minimization to converge to a self-consistent
fit. When the radial velocities have grown to their full observed
values, the code produces a self-consistent $\sqrt{\chi^{2}}=1.82$
fit, which we list in the second column of Table 4. (All of the fits
in table 4 correspond to Epoch JD 2447578.730, the time at which the
first radial velocity observation of the star was made).

With the exception of the eccentricity of the middle planet, which has
dropped from its large value of 0.34, the osculating orbital elements
for the self-consistent fits are quite similar to the summed Kepler
fit. In contrast to the situation with GJ 876, the imposition of
planet-planet interactions has not improved the $\chi^{2}$
statistic. Furthermore, examination of the 3:1 resonance arguments,
$\theta_{1}=3\lambda_{c}-\lambda_{b}-2\varpi_{b}$,
$\theta_{2}=3\lambda_{c}-\lambda_{b}-\varpi_{c}-\varpi_{b}$, and,
$\theta_{3}=3\lambda_{c}-\lambda_{b}-2\varpi_{c}$, show that while the
model systems are close to resonance, none of the resonance arguments
are librating for any of the foregoing fits or for the two additional
fits discussed below.. This is illustrated in Figure 12, where the
time variation of the resonant argument $\theta_{1}$ is plotted for
both the 3-Keplerian fit and for the self-consistent fits. It seems
clear that the 55 Cancri system will likely turn out to be very
interesting dynamically, even if it is not in resonance today.  It
likely was in the resonance in the past, and the fact that it is
currently not indicates an intriguing past, possibly including tidal
dissipation (Greg Novak, personal communication, 2002).

The failure of the Levenberg-Marquardt routine to significantly
improve the $\chi^{2}$ statistic suggests that a true global minimum in
the three-planet parameter space was simply not located by the {\it
locally} convergent Levenberg-Marquardt algorithm. The failure of
Levenberg Marquardt led us to adopt a genetic algorithm (Goldberg
1989) as implemented by D.L. Carroll (1999) for public domain use. The
genetic algorithm starts with a population of osculating orbital
elements, each referenced to the epoch of the first radial velocity
observation. Each set of elements (genomes) describes a unique
four-body integration, and therefore an associated radial velocity
curve for the central star.  The fitness of a particular genome is
measured by the $\chi^{2}$ value of its fit to the radial velocity
data set. At each generation, the genetic algorithm evaluates the
$\chi^{2}$ fit resulting from each parameter set and crossbreeds the
best members of the population to produce a new generation.

Extensive use of the genetic algorithm also failed to find a significantly
improved $\chi^{2}$ value
over that provided by the summed Keplerian fit, in contrast to
the algorithm's excellent performance
on test problems involving strongly interacting planets (Laughlin \& 
Chambers 2002). The best fit which we evolved is listed in the third
column of Table 4. This fit has $\chi^{2}$=1.82, and has osculating
orbital elements which are very similar to the fit obtained by
slowly increasing the strength of the planet-planet interactions
(second column of Table 4).

In summary, we have investigated three numerical strategies for
producing self-consistent three-planet fits to the 55 Cancri radial
velocity data set. All three methods provide fits with $\chi^{2}$
statistics that are essentially equivalent to the best 3-Kepler fit,
and all three fits are in quite good agreement, suggesting that the
system lies just outside of the 3:1 resonance.  We stress, however,
that it is not yet completely clear whether the system is indeed in
the 3:1 resonance. Further dynamical fitting and further observation
of the star will be required to definitively identify the dynamical
relationship between the inner and the middle planet.

\section{Stability of Earth--Mass Planets near 1 AU}

The architecture of the 55 Cancri system, with giant planets in 14.65
and 4680 day orbits, leads to an anthropocentric question: Is a
terrestrial planet stable at 1 AU? The answer is yes. The large period
separation between the middle and outer planets admits a broad zone of
stable orbits in the so-called habitable zone of this system. We have
done a number of representative integrations, which combine the
nominal co-planar three--planet system listed in Tables 3 and 4 with
an Earth-mass planet on an initially circular 1 AU orbit. The orbital
elements are assumed to map onto planetary configuration built in
Jacobi coordinates. Apse precession due to general relativistic
effects is accomodated by augmenting the stellar gravitational
potential with an approximate post-Newtonian correction (see e.g. Saha
\& Tremaine 1992).  This improvement has a qualitatively negligible
effect on the results.

In typical cases, perturbations from the giant planets
cause the eccentricity of the Earth-mass planet to oscillate over a 27,000
year period with an amplitude in $e$ of 0.03 . Note that perturbations from
the other planets in our own solar system cause Earth to experience
chaotic eccentricity oscillations of similar magnitude and duration
(see e.g. Murray and Dermott, 1999).

\eject
\eject

\section{Discussion and Conclusions}

All $\sim$90 previously reported extrasolar planets reside in smaller orbits than
that of Jupiter in our Solar System.  The duration of Doppler searches
for extrasolar planets had not been long enough to capture an entire
orbital period of 12 years for planets at 5 AU.  Indeed all previously known
planets are known to have semimajor axes less than 4 AU, well within Jupiter's
orbital distance.  Moreover, a large majority of the extrasolar planets 
reside in eccentric orbits.  Thus it has remained inappropriate to compare the
extrasolar planets against Jupiter or Saturn in
our Solar System.

The Lick Observatory Doppler planet search began in 1987 and thus now
has the requisite duration to detect planets having orbital periods of
over a decade.  The velocities of 55 Cnc during 13 years can be
explained nearly adequately by two planets orbiting the star.  We had
previously detected the inner planet to 55 Cnc (``55 Cnc b'', Butler
et al. 1997). Its mass ($>$0.9 \mjup) and circular orbit with a radius
of 0.11 AU from the star represents a class of close--in extrasolar
planets, sometimes called ``hot jupiters'', the first member of which
was 51 Peg (Mayor and Queloz 1995).

The velocities of 55 Cnc now reveal strong evidence of an outer planet
at 5.9 AU, previously suspected due to the additional wobble of 55 Cnc
(Marcy \& Butler 1998).  The reality of an outer planet with an
orbital period of 13--15 yr and a minimum mass of 4 \mjup is securely
supported by the velocities.  Remaining velocity residuals with RMS of
13 \ms are caused either by gas motions on the stellar surface or by
additional orbiting bodies.  The best three--planet fits imply a third
planet having \msini = 0.25 \mjup at 0.24 AU in an orbit with $e$=0.3.
But the three--planet models only partially explain the discrepancies
in the two--planet fit.

As the first extrasolar planet discovered that orbits farther than 4
AU from its host star, the outer planet to 55 Cnc (``55 Cnc c'') is
the first one amenable to direct comparison with the Jovian planets in
our Solar System.  The outer planet has an orbital eccentricity of
0.16$\pm$0.06 to be compared with 0.048 and 0.054 for Jupiter and
Saturn in our Solar System.  Thus 55 Cnc c has a modest orbital
eccentricity corresponding to an orbital path that carries it as close
as 5 AU from the star and as far as 6.8 AU.

At a typical angular separation from the star of 0.47 arcsec, the
planet 55 Cnc c will induce an astrometric wobble in the host star
with an amplitude of 1.8 milliarcsec/$\sin i$ relative to the
barycenter.  This gives some hope that Hipparcos, the {\it Hubble
Space Telescope FGS}, or some other ground--based astrometric program
could detect the wobble.  We carried out an analysis of the Hipparcos
and Tycho--2 catalog astrometry similar to that described by Pourbaix
(2001) and Pourbaix \& Arenou (2001).  In some analyses, we used the
long term proper motion from Tycho--2 to search for a residual
astrometric wobble in the Hipparcos astrometry of 55 Cnc.  In other
analyses we searched for a self--consistent solution of all available
astrometry from both Hipparcos and Tycho-2.  We found no significant
wobble in 55 Cnc at a level of $\sim$3 milliarcsec over the time scale of
the life time of Hipparcos, 4 years.  However this time baseline is
too short to place any constraints at all on the inclination of the
orbit of 55 Cnc c.  The wobble of the star caused by it would be
nearly linear during 4 yr, and hence would be absorbed into the
solution of the proper motion of the star.  Similarly, no HST FGS
astrometry has adequate duration.  We are unaware of any ground--based
astrometry that has adequate precision to detect the companion.  Thus we
are not able to place any constraints on the orbital inclination of 55
Cnc c, and hence cannot place an upper limit on its mass.  Both SIM
and GAIA would carry adequate astrometric precision to detect the
motion of the star due to 55 Cnc c.  But a mission lifetime of at least 7--10
years (nearly one orbital period) will be necessary to separate
the proper motion from orbital parameters.  Astrometry having
a precision of $\sim$20 $\mu$as, coupled with velocities, 
would constrain the mass of 55 Cnc c to within a few percent.

The velocity residuals to the two--planet model exhibit an RMS
of 12 \ms and a strong periodicity of 44.3 d (Fig 9).  These residuals are
certainly not due to any instrumental effect, as we are monitoring 300
stars at Lick, including many stars with spectral type G5--K0. 
No periodicities between 40--50 d are seen among those other stars.

The proposition that the 44.3 d period is caused solely by surface
effects on the star seems unlikely.  The stellar characteristics of 55
Cnc (section 2) suggest that it is a quiescent star of age 3--8 Gyr,
showing very little variation in the usual surface diagnostics.  The
photometric variation is no more than 1 mmag, and the level and
activity in the CaII K-line emission reversal is small.  Such stars
are well studied by precision Doppler programs and they exhibit
velocity variations of 2--5 \ms, presumably caused by turbulence and
patchy magnetic regions located non--uniformly over the surface.  Thus
we cannot support a model in which the velocity residuals with RMS of
12 \ms are intrinsic to the stellar surface.

In contrast, our attempts to fit the velocities, notably the 44.3 d
period, with a third planet yielded a significant improvement in the
reduced $\chi^2$ compared with that of a simple two--Keplerian fit.
Neither a triple--Keplerian model nor a triple--planet Newtonian model
succeeded in diminishing the value of $\chi_{\nu}^2$ to a value of
1.0--1.5, but instead left $\sqrt{\chi^2_{\nu}}$=1.8 .  Moreover the
near coincidence in periods between the 44.3 d velocity period and the
35--42 d rotation period leaves us uncomfortable about the
interpretation of the 44.3 d period.  Nonetheless, the chromospheric
and photometric quiescence of the star is not consistent with stellar
surface effects as the cause of the 13 \ms velocity variations.
Indeed, we have never seen such large velocity amplitude and coherence
in such a quiet star.  This issue is examined carefully by Henry et
al. (2002).  Thus we favor the model that includes a third planet with
that period.

Because the value of $\chi^2$ remains too large, even with a model
that contains three planets, one may consider alternative models.
Perhaps 55 Cnc contains yet an additional planet located in the gap
between 0.25 and 5 AU.  The simulations presented here show that a
low--mass planet could persist stably there indefinitely.  Planets of
sub--saturn mass located between 0.25 and 5 AU would be difficult to
detect securely but would cause velocity variations of a few \ms, as
seen in our velocity measurements.

Another possibility is that rotational modulation of surface
inhomogeneities is stronger in metal--rich stars than is seen in
solar--metallicity stars.  In that case, the 44.3 d period could be
caused by stellar rotation after all.  Such a hypothesis requires that
surface effects both cause a stellar ``jitter'' of 13 \ms and remain
coherent in phase over time scales of years.  This could occur if one
longitude maintains its inhomogeneity (spot, magnetic field) for a
duration of years.

We also note that the lack of a dust disk (Jayawardhana et al.~2002,
Schneider 2001) provides limits to the evolution of debris
disks in the face of a giant planet at Jovian distances.  One wonders
whether such Jupiter analogs tend to enhance to production of dust via
enhanced collision rates between the comets and asteroids or instead promote
the clearing of the dust due to gravitational perturbations during the
lifetime of the star.

The separation of 0.47 arcsec between the star and the outer planet, 
makes this system a likely target for future coronographic imaging and
interferometric nulling, especially from spaceborn telescopes.
The outer planet, ``55 Cnc d'', subtends a fraction of the
sky, $f = 1.6\times10^{-9}$, as seen from the star. 
The wavelength--dependent albedo of giant planets in general is under
active investigation (Seager \& Sasselov 1998; Marley et al. 1999;
Goukenleuque et al. 2000; Sudarsky, Burrows, \& Pinto 2000).  The
albedo at visible wavelengths is likely to be $\sim$1/2 .  Thus,
one expects the planet, 55 Cnc d, to be fainter than the host star
by a factor of $0.8\times10^{-9}$ at optical wavelengths.  
This implies a contrast of 22.7 mag at V band, 
and an apparent magnitude, $V$ = 28.7, for the planet.

With its high abundance of Fe, C, Si, and other heavy elements, along
with its age of $\sim$5 Gyr,  the 55 Cnc system makes an
intriguing target for questions of organic chemistry and biology.  
A rocky planet at roughly 1 AU remains a viable prospect
dynamically.  Moreover the inner two planets and the outer planet 
presumably formed from protoplanetary disk material.
These extent planets beg the question of
the final repository of the disk material that presumably
existed between 0.3 and 5 AU.  The migration of the inner two
planets would not have cleared the region at 1 AU.  Indeed, such
migration could have occurred by virtue of the two planets
delivering angular momentum and energy to the disk material
outward of their orbits.  Indeed, the presence of two
planets at 0.1 and 0.24 AU suggests that material
existed between 0.3 and 5 AU, serving as the recipient of
their orbital angular momentum.

We expect to detect, within the next 5 years, a sizable population of
Jupiter--mass planets orbiting at 4--6 AU.  These planets may serve as
signposts of planetary systems characterized by architectures similar
to that of our Solar System: gas giants beyond 5 AU and rocky planets
closer in.  Such jupiters are amenable to direct comparison
with Jovian planets in our Solar System and will permit
characterization of the properties of planetary systems in general.

\eject
\eject

\acknowledgements

We thank Chris McCarthy, Eric Williams, Amy Reines, David Nidever,
Heather Hauser, Eric Nielsen, Bernie Walp, and Mario Savio for taking
spectra of 55 Cancri at Lick Observatory.  We thank Greg Novak for
for dynamical calculations of the three--planet system and Jason
T. Wright for chromospheric and spectroscopic measurements.
We thank Jeff Valenti for many stimulating conversations.
We thank Tony Misch, Wayne Earthman, Keith Baker, and Jim Burrous
for tuning the instrumentation at Lick Observatory 
to unusually fine standards.
We acknowledge support by
NSF grant AST-9988358 (to SSV), NSF grant AST-9988087 and travel
support from the Carnegie Institution of Washington (to RPB), NASA
grant NAG5-8299 and NSF grant AST95-20443 (to GWM), and by Sun
Microsystems.  
We also acknowledge NASA grant NCC5-511 and NSF
grant HRD-9706268 (to G.Henry).
We thank the UCO/LICK Research Unit of the
University of California, and especially Directors Dr. Robert Kraft
and Dr. Joseph Miller for their foresightful allocations of Lick
Observatory telescope time.  This research has made use of the Simbad
database, operated at CDS, Strasbourg, France.

\clearpage

\begin{figure}
\centerline{\scalebox{.75}{\rotatebox{90}{\includegraphics{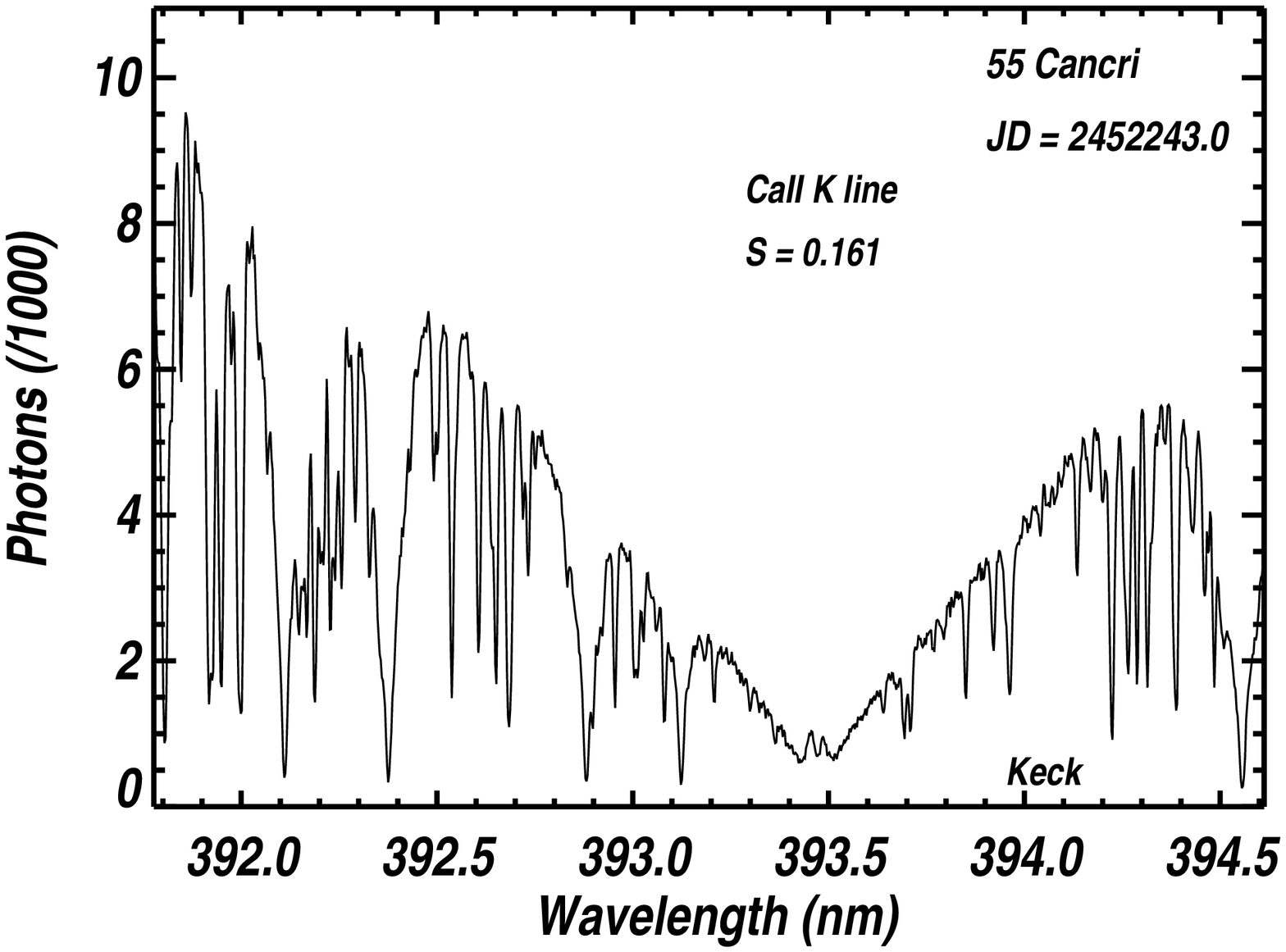}}}}

\caption{Spectrum of 55 Cancri near the Ca II K line showing the 
modest central emission reversal in the core.  The star is relatively quiescent,
typical for a middle--aged star, with a Mt. Wilson 
chromospheric index, S = 0.16.  The deep absorption lines
visually reveal the high metallicity, found here to be [Fe/H]=+0.27 .
This metallicity along with its luminosity and T$_{\rm eff}$ implies a stellar
mass of 0.95 \msun.} 
\label{hkplot} 
\end{figure}

\begin{figure} 
\centerline{\scalebox{.75}{\includegraphics{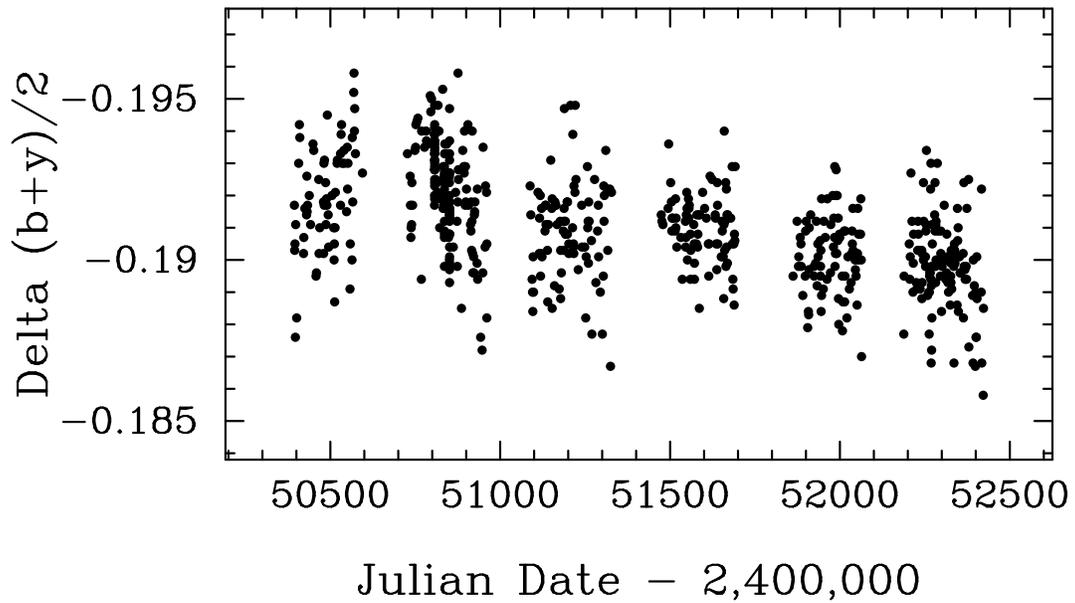}}}
\caption{Differential Str\"{o}mgren photometry of 55 Cnc during the past six observing 
seasons acquired with a 0.8~m APT.  The night-to-night RMS is $~\sim$1.2 mmag 
within each season, consistent with the usual measurement precision.  A 
downward trend of 2 mmag over the 6 years indicates a modest magnetic cycle, 
typical for a middle-aged G8-type main sequence star.}
\label{cor_lin} 
\end{figure} 

\begin{figure} 
\centerline{\scalebox{.75}{\rotatebox{0}{\includegraphics{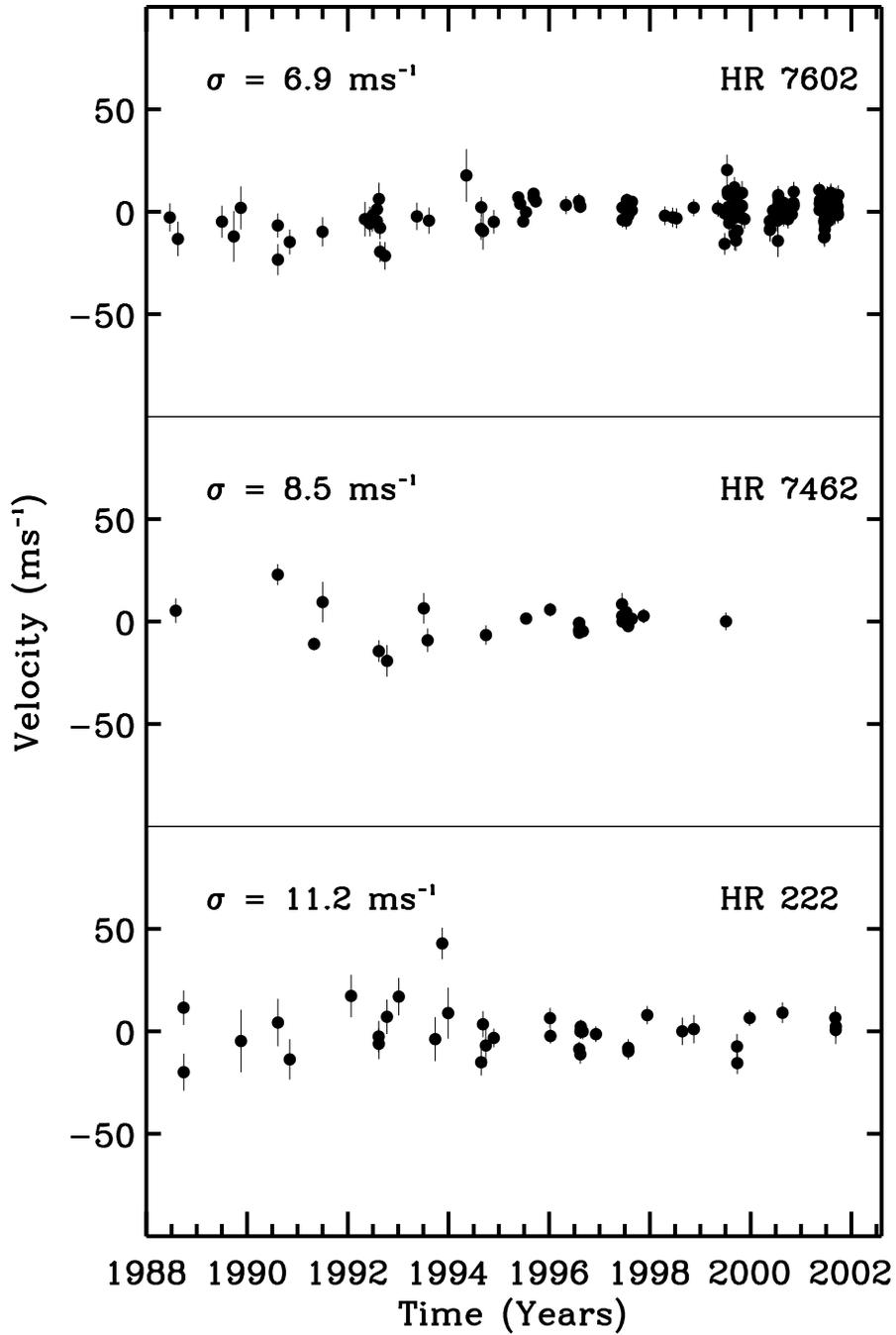}}}}
\caption{Comparison Stars.  Doppler shift measurements for three different
GK dwarfs observed from 1989 to present taken with the same
instrumental setup and the Lick telescopes.  These stars demonstrate
Doppler precision and long--term stability at a level of
10 \ms.}
\label{stable_stars} 
\end{figure} 

\begin{figure} 
\centerline{\scalebox{.75}{\rotatebox{90}{\includegraphics{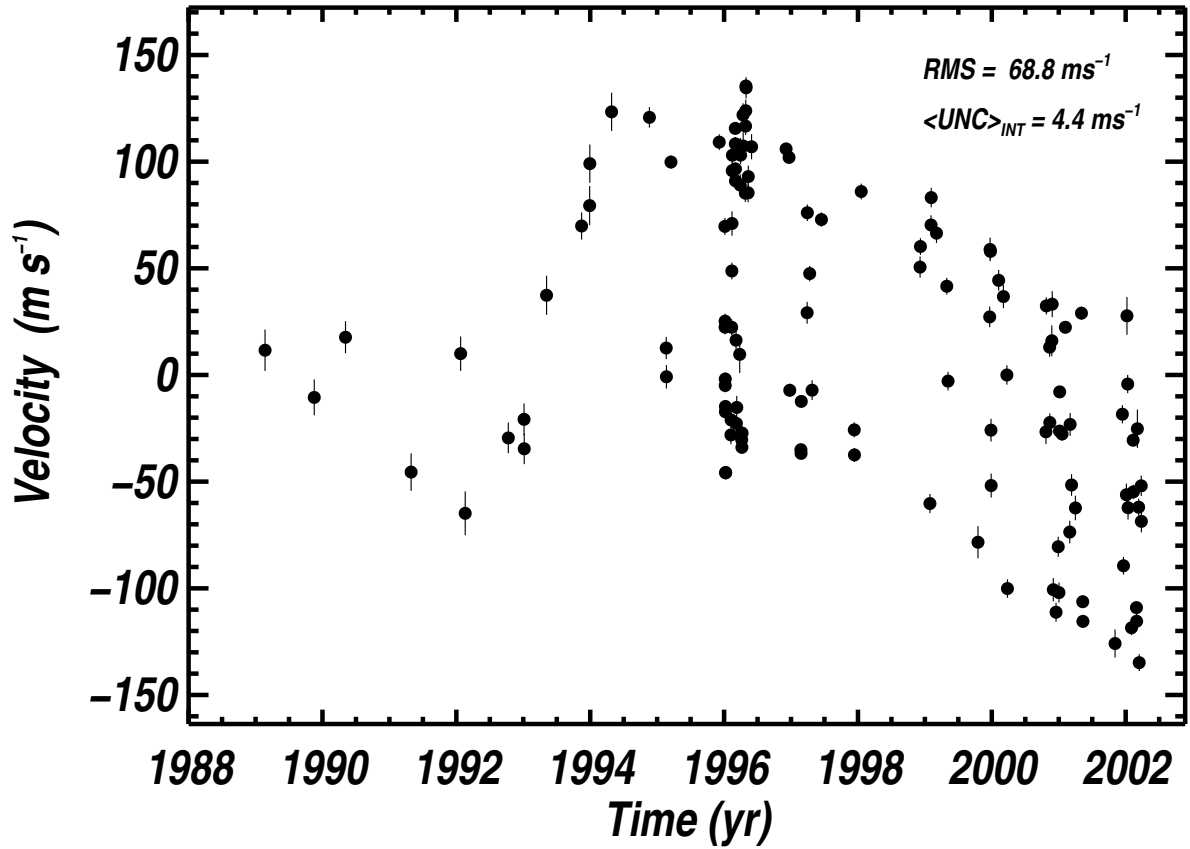}}}}
\caption{Doppler shift measurements for 55 Cancri from 1988 to 2002.
Precision improved from $\sim$10 \ms (1988--1994) to $\sim$4 \ms
(1995--2002).  The 14--year time scale of velocity variations is visible to
the eye, along with the short--period variations caused by the
14.65--d period planet.}
\label{raw_vel} 
\end{figure} 
 
\begin{figure} 
\centerline{\scalebox{.75}{\rotatebox{90}{\includegraphics{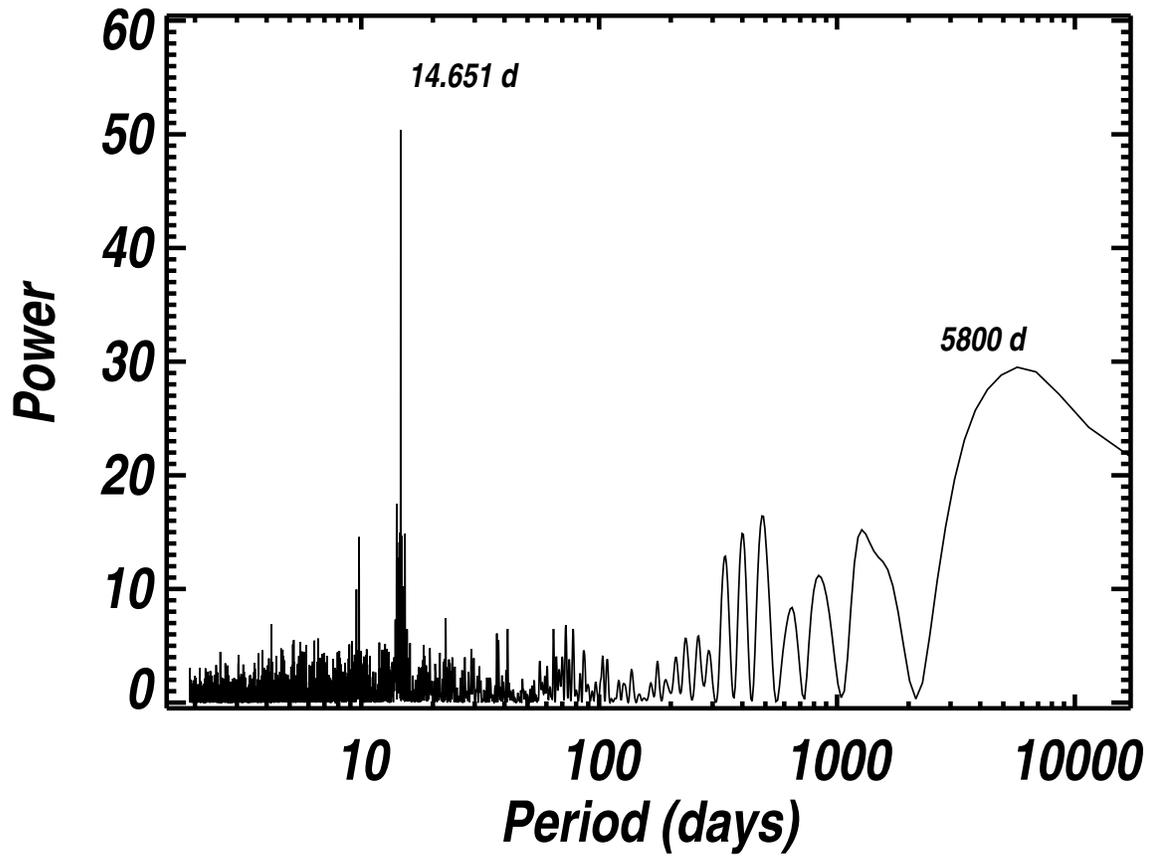}}}}
\caption{Periodogram of the all of the Doppler shift measurements
of 55 Cancri (shown in Figure \ref{raw_vel}).  The tallest two peaks are
both statistically significant at $P$=14.65 d and $P$ = 5800 d.}
\label{periodogram} 
\end{figure} 
 
\begin{figure} 
\centerline{\scalebox{.75}{\rotatebox{90}{\includegraphics{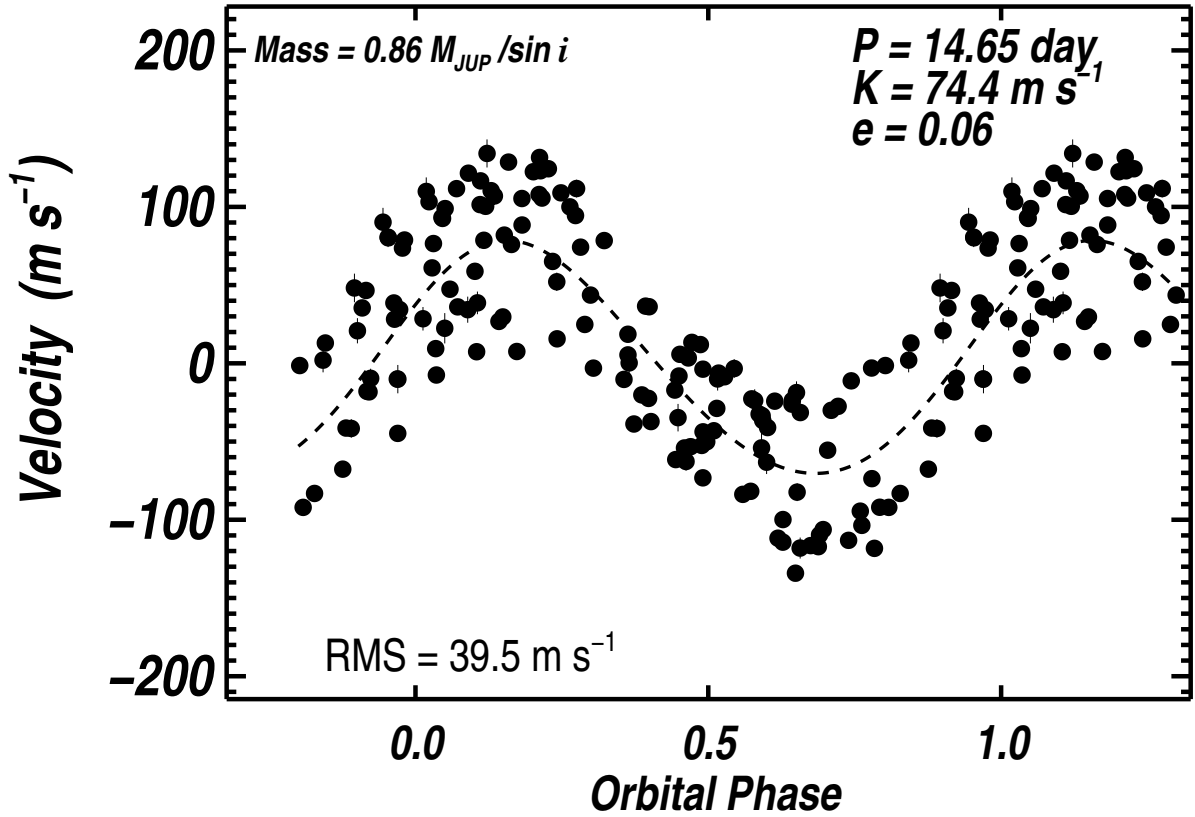}}}}
\caption{Velocities for 55 Cancri, phased at the best--fit period of
14.65 d which corresponds to the tallest peak in the periodogram (Figure \ref{periodogram}).
The dashed line represents the best--fit (single--planet) keplerian.
The RMS of 38 \ms is well above the typical velocity errors of 3--10 \ms,
showing the inadequacy of a single--planet model.} 
\label{oneplanet} 
\end{figure}

\begin{figure} 
\centerline{\scalebox{.75}{\rotatebox{90}{\includegraphics{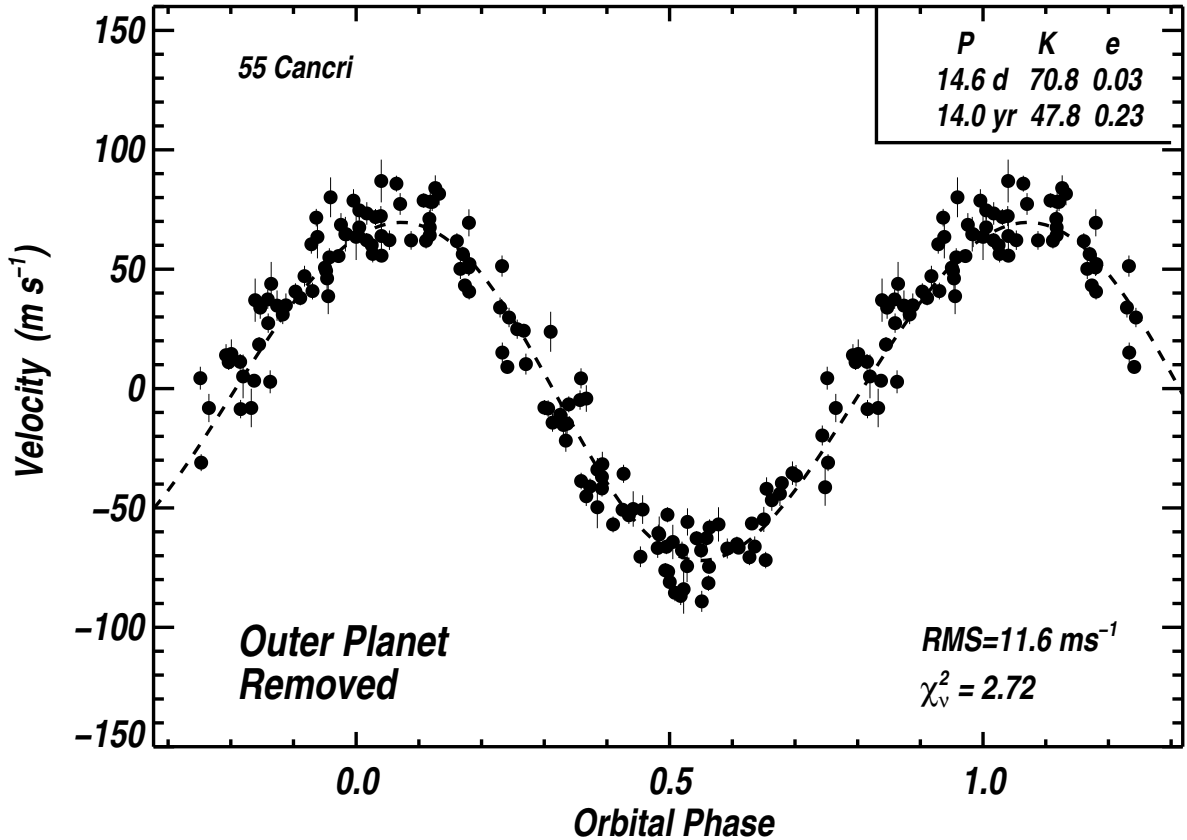}}}}
\caption{A Two--Keplerian model (solid line) is fit to the measured velocities.
Here the velocity wobble caused by the outer planet has been removed for clarity,
leaving the wobble caused by the inner planet.
These residual velocities (dots) are phased at the 14.65--d period
of the inner planet, showing the significant reduction in the
residuals compared with Fig \ref{oneplanet} . By including a
second planet in the model, the RMS of the residuals dropped from 39.5
\ms to 11.6 \ms.}
\label{two_planet_rem_c} 
\end{figure} 

\begin{figure} 
\centerline{\scalebox{.75}{\rotatebox{90}{\includegraphics{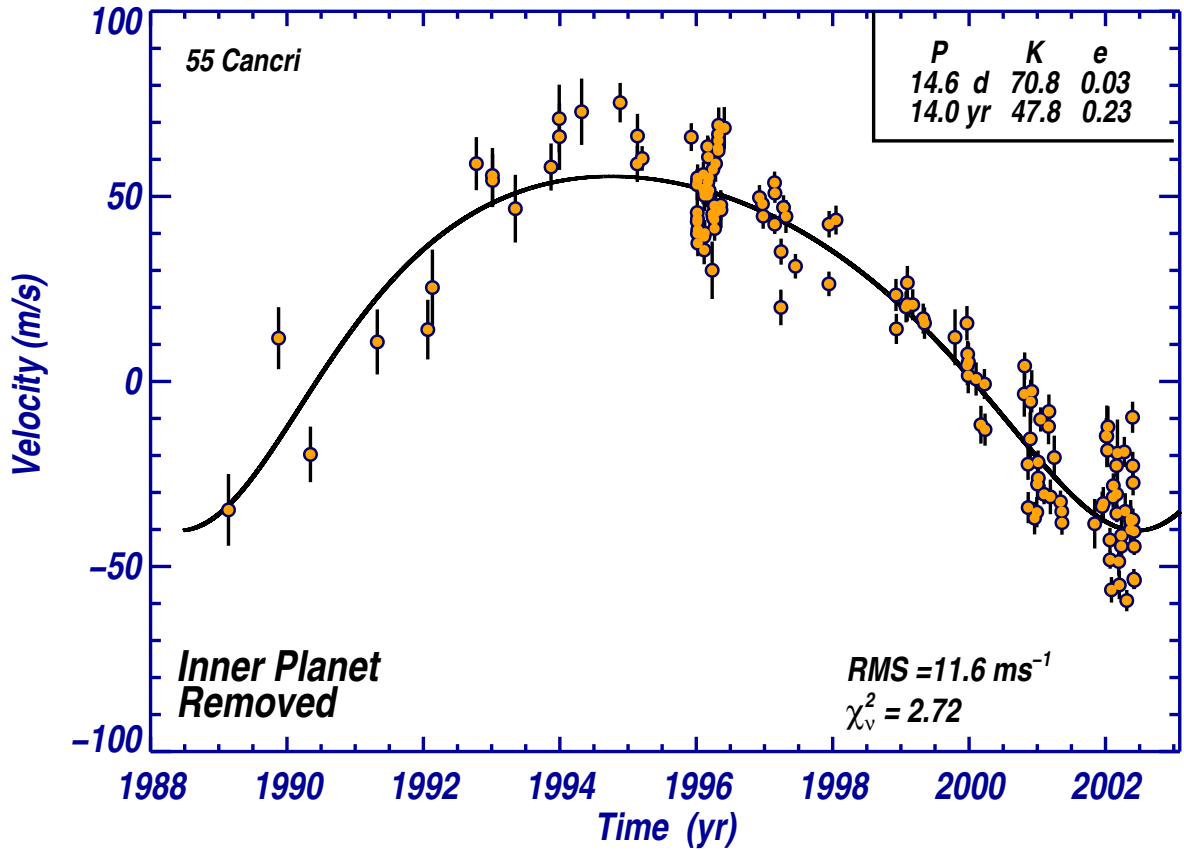}}}}
\caption{A Two--Keplerian model (solid line) is fit to the measured
velocities (as in Fig \ref{two_planet_rem_c}).  Here the velocity wobble caused by the
inner planet ($P$=14.65 d) has been removed.  The residual velocities (dots) are
plotted versus time, and the best--fit Keplerian motion of the outer
planet has an orbital period of 14 yr.  By including a second planet
in the model, the RMS of the residuals dropped from 38.5 \ms to 11.6 \ms.}
\label{two_planet_rem_b} 
\end{figure}

\begin{figure} 
\centerline{\scalebox{.75}{\rotatebox{90}{\includegraphics{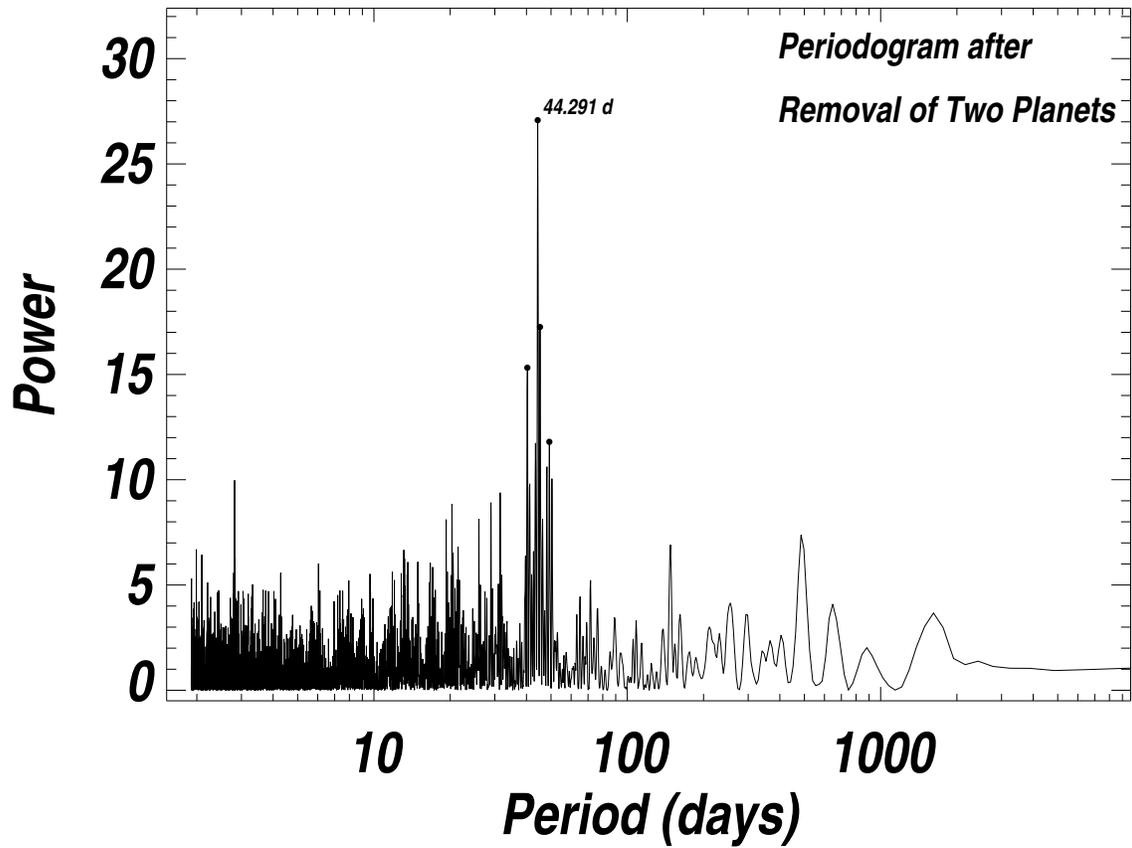}}}}
\caption{Periodogram of the velocity residuals that are left when the
predicted velocities of the double--Keplerian model are subtracted
from the original velocities.  The peak at period 44.3 d has a false alarm
probability under 0.1\%, implying that the periodicity is real.  The
cause could be a third planet or permanent inhomogeneities of the velocity 
field on the stellar surface.  The latter is unlikely because 55 Cnc is a 
quiet star.}
\label{periodogram_rem2} 
\end{figure} 

\begin{figure} 
\centerline{\scalebox{.35}{\rotatebox{90}{\includegraphics{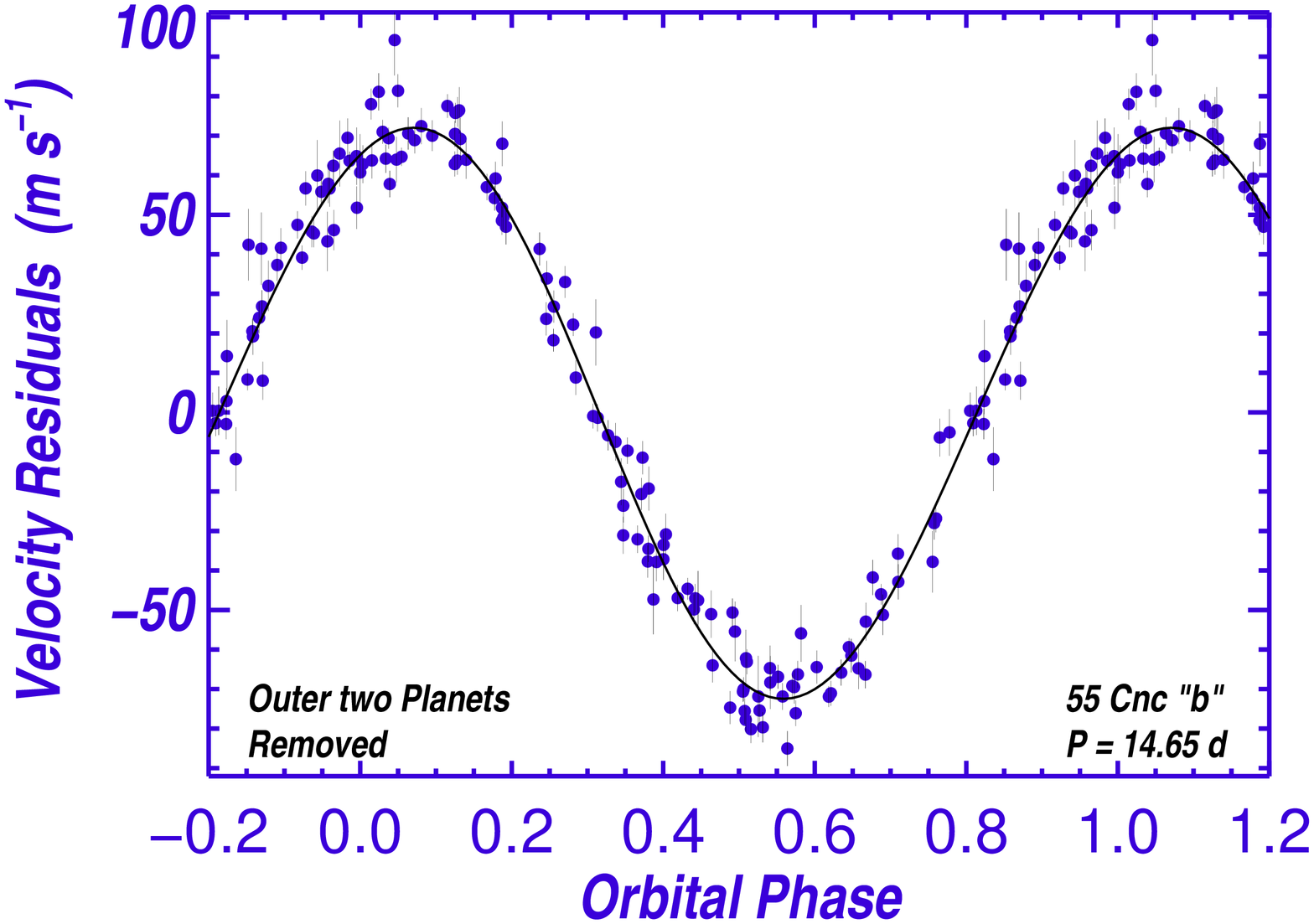}}}}
\bigskip
\centerline{\scalebox{.35}{\rotatebox{90}{\includegraphics{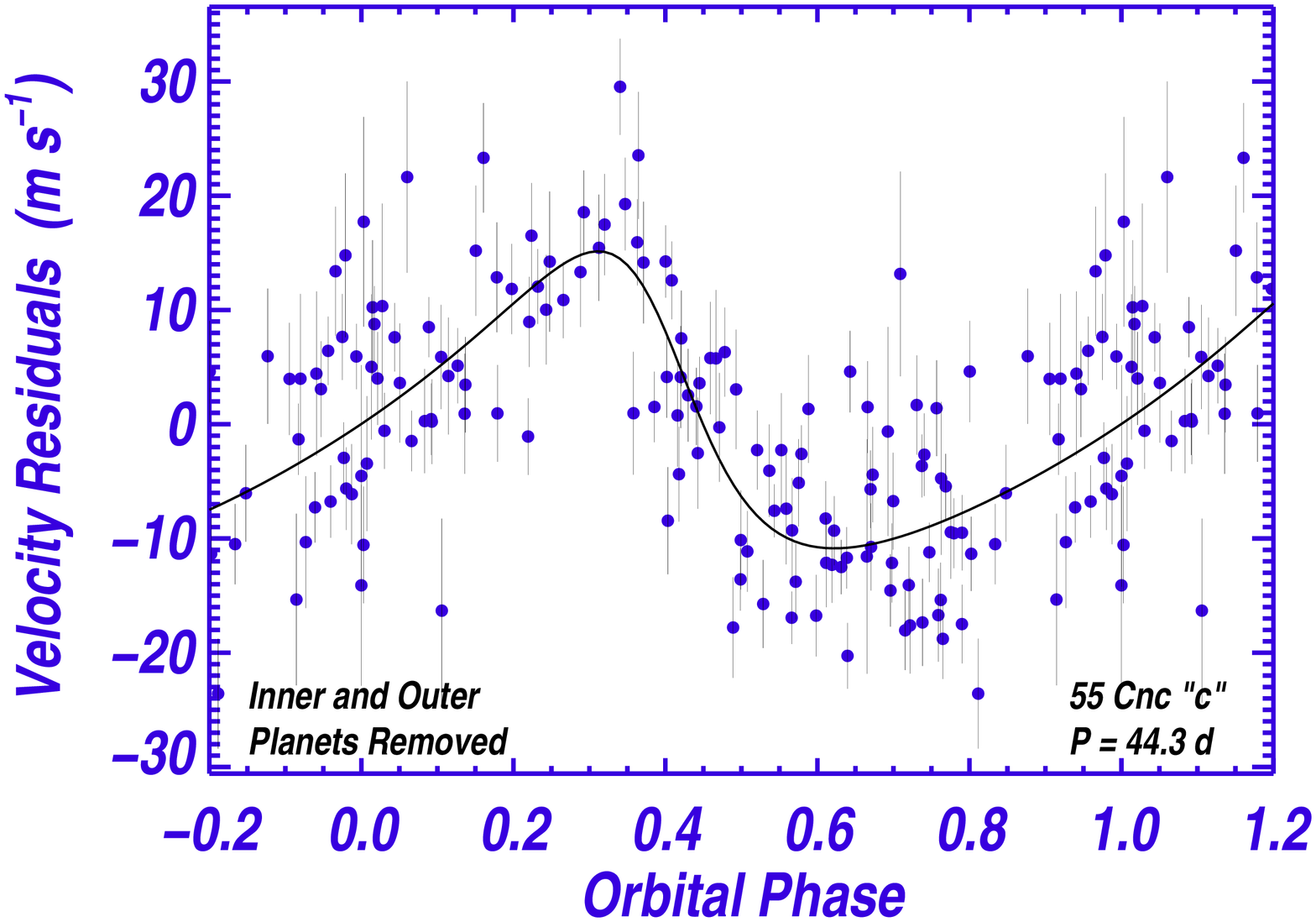}}}}
\bigskip
\centerline{\scalebox{.35}{\rotatebox{90}{\includegraphics{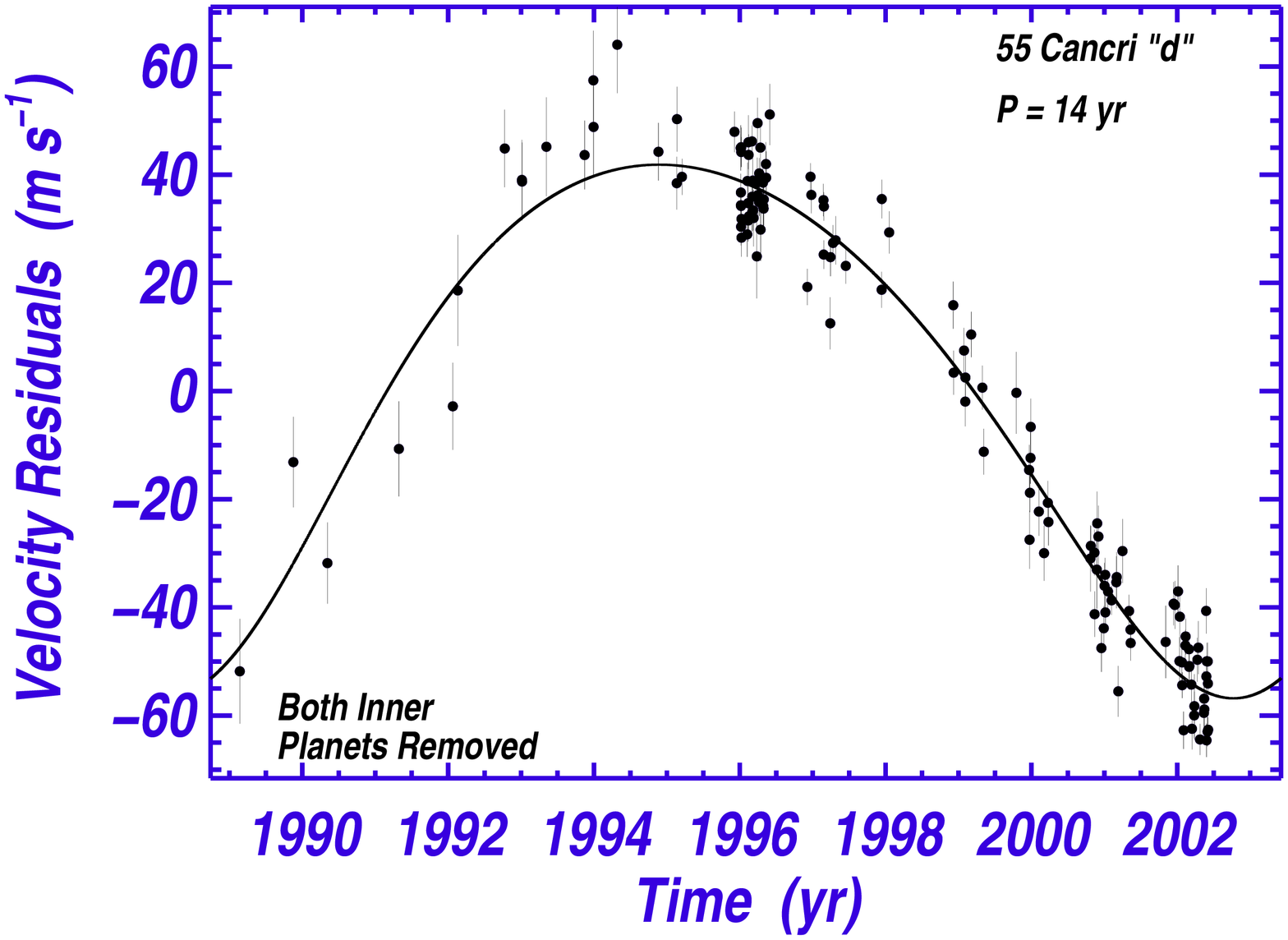}}}}
\caption{A three--Keplerian orbital fit to the velocities for 55 Cnc.
The velocities and fits for each of the three planets are shown separately for clarity,
by subtracting the effects of the other two planets. The panels contain
(top) inner planet ``b'', (middle) middle planet ``c'', and (bottom) 
outer planet ``d''.}
\label{three_kepler_fit} 
\end{figure} 

\begin{figure} 
\centerline{\scalebox{.7}{\rotatebox{0}{\includegraphics{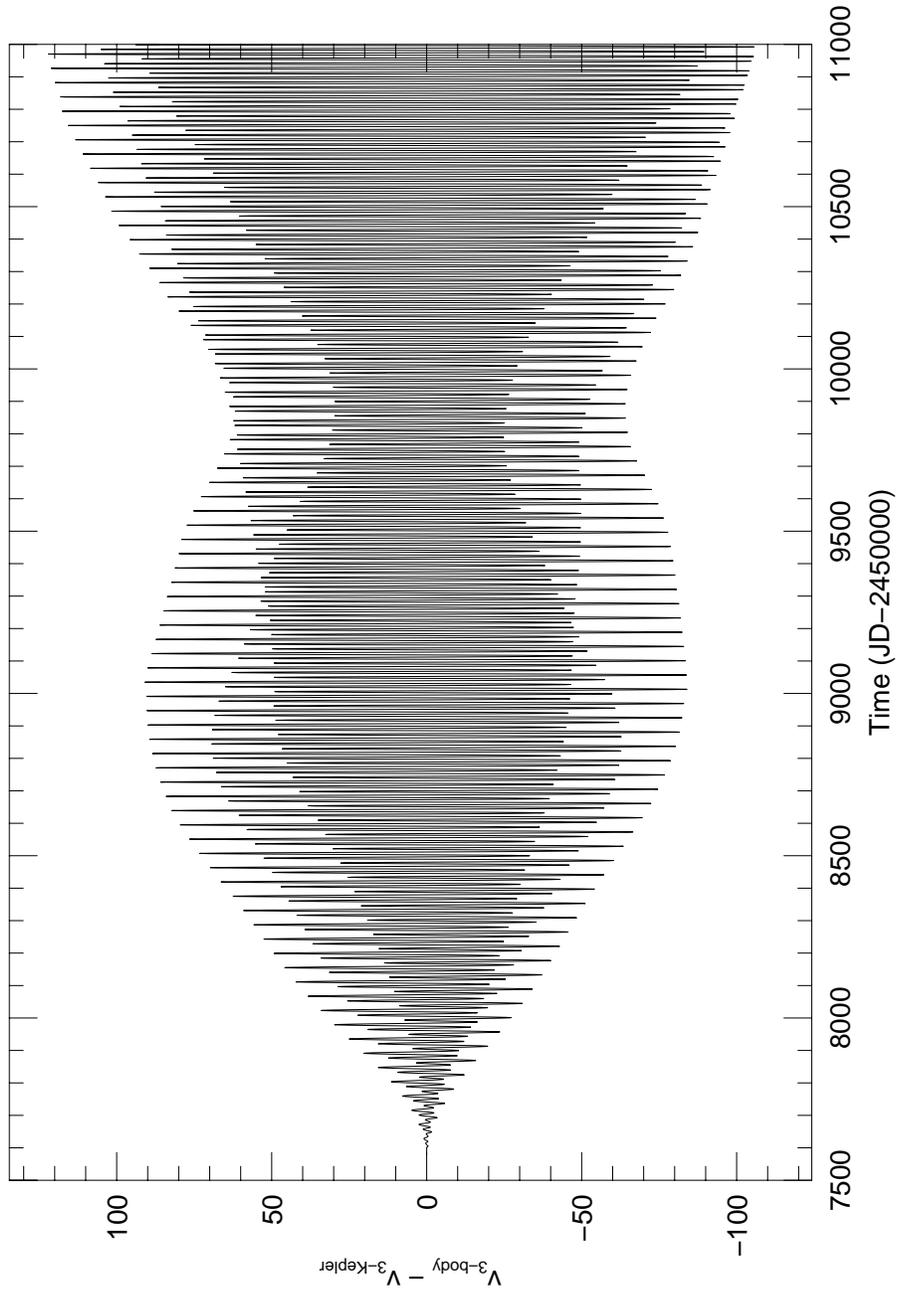}}}}
\caption{ Running difference between the radial velocity curve of the
star under the influence of summed Keplerian motion, and under the
full four--body Newtonian motion.  Here three planets are assumed with semimajor axes
of 0.11, 0.24, and 5.5 AU, the middle planet remaining hypothetical with period
of 44.3 d.    The two simulations of stellar
motion differ by 100 \ms after 10 years, indicating that a
self--consistent Newtonian fit wil be required if the middle planet
with period 44 d actually exists.}
\label{running_diff} 
\end{figure}

\begin{figure} 
\centerline{\scalebox{.75}{\rotatebox{90}{\includegraphics{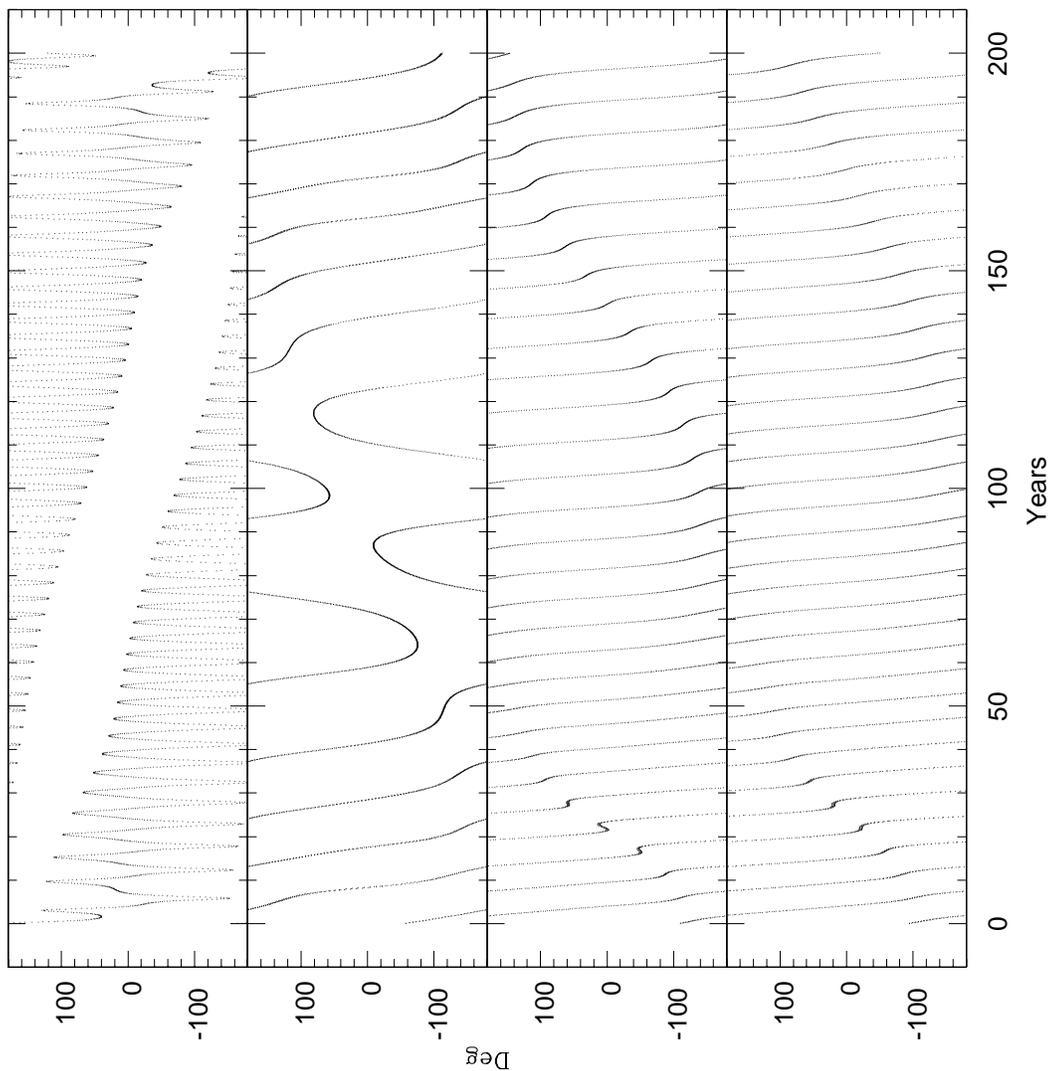}}}}
\caption{ Time behavior of the 3:1 resonant argument
$\theta_{1}=3\lambda_{c}-\lambda_{b}-2\varpi_{b}$, for {\it Top Panel}
summed Keplerian fit, {\it Second Panel} self-consistent 4-body fit
\#1 (see Table 4), {\it Third Panel} self-consistent 4-body fit \#2,
{\it Bottom Panel} self--consistent 4-body fit \#3.  Resonance
behavior is not quite achieved in any of these simulations.}
\label{resonance} 
\end{figure}

\clearpage

\begin{deluxetable}{rrlllllllll}
\tablenum{1}
\tablecaption{Stellar Properties}
\label{candid}
\tablewidth{0pt}
\tablehead{
\colhead{Star} &  \colhead{Spec} &  \colhead{V} & \colhead{B-V} & \colhead{M$_{\rm Star}$} & \colhead{log(R'$_{\rm HK}$)} & \colhead{[Fe/H]} & \colhead{dist} & P$_{\rm Rot}$ & Age\\
\colhead{} & \colhead{type} & \colhead{(M$_{\odot}$)} & \colhead{(mag)}  &  &  &  & (pc) & (d) & (Gyr)
}
\startdata
 55 Cnc  & G8 ~V &  5.95 & 0.87 & 0.95 & -5.02 & +0.27 & 12.53 & 35--42 & 3--8 \\
\enddata
\end{deluxetable}

\clearpage

\begin{deluxetable}{rrr}
\tablenum{2}
\tablecaption{Measured Velocities for 55 Cancri}
\label{measuredvels}
\tablewidth{0pt}
\tablehead{
JD$~~$ & RV$~~$ & Unc.$~~$ \\
(-2450000)   &  (m s$^{-1}$) & (m s$^{-1}$)
}
\startdata
\tableline
-2421.2700 &      26.76 &   9.7  \\
-2152.9563 &       4.64 &   8.4  \\
-1982.3118 &      32.84 &   7.5  \\
-1624.3308 &     -30.37 &   8.8  \\
-1353.9989 &      25.15 &   8.1  \\
-1329.2395 &     -49.69 &  10.3  \\
-1093.9639 &     -14.35 &   7.2  \\
-1008.0364 &      -5.64 &   7.4  \\
-1007.1095 &     -19.44 &   7.2  \\
 -885.2714 &      52.52 &   9.2  \\
 -693.9788 &      84.99 &   6.4  \\
 -650.1551 &      94.56 &   9.2  \\
 -649.0730 &     114.22 &   9.0  \\
 -530.3522 &     138.50 &   9.0  \\
 -323.9368 &     135.92 &   5.3  \\
 -232.2260 &      17.73 &   4.9  \\
 -231.1678 &       1.01 &   6.0  \\
 -206.2017 &     113.31 &   3.3  \\
   56.9882 &     127.44 &   3.8  \\
   87.8823 &      82.81 &   4.2  \\
   88.9186 &      40.90 &   3.2  \\
   89.0055 &      40.50 &   3.5  \\
   89.7764 &      10.12 &   3.5  \\
   89.9859 &       7.80 &   3.3  \\
   90.7448 &      -1.91 &   2.8  \\
   90.8930 &      -4.24 &   3.6  \\
   91.8485 &     -32.21 &   3.3  \\
   91.9696 &     -36.71 &   3.5  \\
  120.8739 &     -18.34 &   4.2  \\
  121.8886 &     -19.37 &   3.4  \\
  124.8562 &      17.33 &   3.5  \\
  125.7787 &      39.75 &   3.9  \\
  126.8369 &      83.19 &   5.0  \\
  127.8525 &     103.26 &   3.8  \\
  128.8662 &     104.70 &   3.6  \\
  144.7105 &     126.81 &   3.0  \\
  144.8541 &     112.26 &   3.2  \\
  145.6248 &     104.37 &   3.1  \\
  145.7650 &      98.80 &   3.6  \\
  148.8883 &      16.32 &   5.2  \\
  150.7493 &     -19.83 &   3.7  \\
  152.6622 &      -6.85 &   5.3  \\
  168.7494 &       6.37 &   7.8  \\
  171.7381 &      96.98 &   4.6  \\
  172.6893 &     120.98 &   4.7  \\
  173.7246 &     109.68 &   3.9  \\
  179.7331 &     -29.13 &   2.6  \\
  180.6884 &     -27.12 &   2.8  \\
  181.6339 &     -23.14 &   3.3  \\
  186.7391 &     116.00 &   4.9  \\
  187.6855 &     111.31 &   5.8  \\
  199.6818 &      84.56 &   4.0  \\
  200.7012 &     107.64 &   3.9  \\
  201.6854 &     125.87 &   4.8  \\
  202.6895 &     132.98 &   3.3  \\
  203.6860 &     128.73 &   4.8  \\
  214.6864 &      78.00 &   4.1  \\
  215.6724 &      97.59 &   4.9  \\
  233.6917 &     116.06 &   5.7  \\
  422.0056 &     114.70 &   3.4  \\
  437.9283 &     109.99 &   2.5  \\
  441.9539 &       0.61 &   3.4  \\
  502.7805 &     -21.74 &   3.0  \\
  503.7636 &     -25.69 &   2.6  \\
  504.7734 &       1.33 &   2.6  \\
  536.7737 &      42.94 &   4.8  \\
  537.7634 &      80.90 &   3.5  \\
  550.7260 &      50.93 &   3.3  \\
  563.7190 &       2.92 &   4.5  \\
  614.6935 &      78.60 &   3.3  \\
  793.9024 &     -24.45 &   3.3  \\
  794.9620 &     -28.09 &   3.6  \\
  831.9320 &     105.80 &   3.9  \\
 1153.0331 &      65.36 &   4.3  \\
 1155.0185 &      80.33 &   4.1  \\
 1206.8777 &     -51.20 &   4.2  \\
 1212.9279 &      83.00 &   4.6  \\
 1213.8834 &      92.76 &   4.6  \\
 1242.7398 &      86.31 &   4.2  \\
 1298.7216 &      38.69 &   4.1  \\
 1305.7085 &      -3.76 &   4.2  \\
 1469.0528 &     -58.87 &   7.6  \\
 1532.9958 &      32.55 &   4.6  \\
 1535.0066 &      63.16 &   5.4  \\
 1536.9490 &      69.49 &   3.6  \\
 1540.0076 &     -12.75 &   4.8  \\
 1540.9832 &     -38.77 &   5.2  \\
 1581.8477 &      47.97 &   4.5  \\
 1607.8268 &      40.42 &   5.1  \\
 1626.7339 &      22.93 &   4.1  \\
 1629.8053 &     -77.45 &   4.3  \\
 1840.0491 &      -5.25 &   6.1  \\
 1842.0338 &      51.73 &   3.7  \\
 1860.0563 &      29.20 &   4.2  \\
 1861.0366 &      -5.93 &   4.2  \\
 1872.0209 &      42.99 &   7.2  \\
 1874.0057 &      56.46 &   5.9  \\
 1880.0176 &     -77.99 &   5.7  \\
 1895.0062 &    -112.16 &   4.4  \\
 1906.9604 &     -68.84 &   4.1  \\
 1910.8976 &     -90.13 &   4.2  \\
 1913.9658 &      -5.69 &   3.1  \\
 1914.9274 &      13.74 &   3.1  \\
 1927.9088 &     -13.92 &   3.3  \\
 1945.9059 &      34.01 &   2.7  \\
 1969.7891 &     -69.37 &   4.8  \\
 1971.8079 &     -13.53 &   4.6  \\
 1979.7506 &     -49.76 &   4.7  \\
 2000.7152 &     -37.28 &   5.9  \\
 2033.7100 &      31.16 &   3.0  \\
 2040.6905 &    -107.42 &   3.1  \\
 2041.6995 &    -112.83 &   3.3  \\
 2217.0452 &    -113.73 &   6.7  \\
 2257.0349 &     -15.85 &   4.0  \\
 2262.9867 &     -87.66 &   4.4  \\
 2278.9359 &     -37.02 &   4.8  \\
 2281.9804 &      38.65 &   8.3  \\
 2285.9884 &       9.73 &   4.4  \\
 2287.9583 &     -45.80 &   5.6  \\
 2298.8652 &      20.02 &   3.3  \\
 2299.7744 &       1.28 &   2.4  \\
 2306.8041 &    -113.92 &   3.4  \\
 2315.8035 &     -18.08 &   2.9  \\
 2316.8553 &     -48.93 &   3.3  \\
 2333.8160 &     -95.52 &   3.2  \\
 2334.7307 &    -105.32 &   4.2  \\
 2335.7892 &     -99.22 &   2.3  \\
 2338.8319 &      -5.75 &   9.1  \\
 2345.7838 &     -57.16 &   3.8  \\
 2348.7789 &    -129.80 &   3.8  \\
 2359.8143 &     -32.88 &   4.4  \\
 2360.6916 &     -58.39 &   5.9  \\
 2375.7349 &     -48.14 &   4.1  \\
 2380.6987 &     -78.82 &   4.9  \\
 2388.6887 &     -34.42 &   2.9  \\
 2408.7083 &    -108.79 &   3.1  \\
 2409.7136 &     -87.73 &   5.1  \\
 2410.7044 &     -63.31 &   4.2  \\
 2419.7174 &     -39.43 &   4.2  \\
 2420.7126 &     -79.41 &   3.8  \\
 2421.7137 &    -109.96 &   3.1  \\
 2422.7224 &    -101.92 &   3.4  \\
 2426.7305 &     -40.44 &   2.7  \\
 2427.6965 &      -3.12 &   3.5  \\
 2428.7037 &      11.76 &   2.3  \\
 2429.7090 &      11.91 &   2.3  \\
\enddata
\end{deluxetable}

\clearpage

\begin{deluxetable}{lccc}
\tablenum{3}
\tablecaption{Triple--Keplerian Orbital Parameters for 55 Cancri}
\label{orbit}
\tablewidth{0pt}
\tablehead{
\colhead{Parameter} & \colhead{Inner ``b''} &  \colhead{Middle ``c''} & \colhead{Outer ``d''} }
\startdata
Orbital period $P$ (d)          &  14.653 (0.0006)  & 44.276 (0.021) & 5360 (400) \\
Velocity amp. $K$ (m\,s$^{-1}$) &  72.2 (1.0)       & 13.0 (1.3)   & 49.3 (2.5)      \\
Eccentricity $e$                &  0.020 (0.02)     & 0.339 (0.21)  & 0.16 (0.06)   \\
$\omega$ (deg)                  &  99.  (35)        & 61 (25)    & 201 (22)       \\
Periastron Time (JD)            & 2450001.479 (-) & 2450031.4 (2.5) & 2452785 (250) \\
a$_1 \sin i$ (AU)               & 9.8$\times10^{-5}$ (10$^{-6}$) & 5.1$\times10^{-5}$ (1$\times10^{-5}$  & 0.0185 (0.002)     \\
Msini (\mjup)                   & 0.84 (0.07)        & 0.21 (0.04)    & 4.05 (0.4)   \\
a (AU)                          & 0.115 (0.003)     & 0.241 (0.005)       &  5.9 (0.9)   \\
\enddata
\end{deluxetable}

\eject

\begin{deluxetable}{lrrr}
\tablenum{4}
\tablecaption{Three--Planet Newtonian Models}
\label{three-planet-newt}
\tablewidth{0pt}
\tablehead{
\colhead{Param}   &  \colhead{4-Body Fit \#1} & \colhead{4-Body Fit \#2} & \colhead{4-Body Fit \#3} }
\startdata
$P_b$ (days)      & 14.653   & 14.654   & 14.653  \\
$P_c$ (days)      & 44.188   & 44.241   & 44.284  \\
$P_d$ (days)      & 5592.09  & 5514.33  & 5483.70  \\
\\
$Mean Anom_b$ (deg)       & 247.065  & 236.105  & 229.38 \\
$Mean Anom_c$ (deg)       & 336.783  & 215.429  & 217.71 \\
$Mean Anom_d$ (deg)       & 5.3461   & 13.408  &  9.407 \\
\\
$e_b$             & 0.013    & 0.039    & 0.031   \\
$e_c$             & 0.080    & 0.158    & 0.142   \\
$e_d$             & 0.146    & 0.150    & 0.150   \\
\\
$\varpi_b$ (deg)  &  93.12   & 104.17   &  109.85  \\
$\varpi_c$ (deg)  & 299.62   & 51.17   & 57.77  \\
$\varpi_d$ (deg)  & 211.67   & 202.79   & 205.11  \\
\\
 Mass$_b$ ($M_{JUP}$) & 0.831    & 0.836    & 0.837   \\
 Mass$_c$ ($M_{JUP}$) & 0.204    & 0.202    & 0.201   \\
 Mass$_d$ ($M_{JUP}$) & 4.363    & 4.192    & 4.189   \\
\\
$\sqrt{\chi^2}$ & 1.85 & 1.82 & 1.82 \\
epoch (JD)  & 2447578.730 & 2447578.730 & 2447578.730 \\
$v_{{\rm epoch}} {\rm offset}\ {\rm m/s}$ & 3.271 & 4.038 & 3.708 \

\enddata
\end{deluxetable}

\end{document}